\documentclass[5p,authoryear]{elsarticle}
\makeatletter
\def\ps@pprintTitle{%
 \let\@oddhead\@empty
 \let\@evenhead\@empty
 \def\@oddfoot{}%
 \let\@evenfoot\@oddfoot}
\makeatother

\usepackage{natbib}
\usepackage{amssymb}
\usepackage{amsmath}
\usepackage{lineno}
\usepackage{hyperref} %[draft]
\usepackage{enumitem}
\usepackage{titlesec}

\usepackage{fancyvrb}
\usepackage{color}
\usepackage{verbatim}

\makeatletter

\def\PY@reset{\let\PY@it=\relax \let\PY@bf=\relax%
    \let\PY@ul=\relax \let\PY@tc=\relax%
    \let\PY@bc=\relax \let\PY@ff=\relax}
\def\PY@tok#1{\csname PY@tok@#1\endcsname}
\def\PY@toks#1+{\ifx\relax#1\empty\else%
    \PY@tok{#1}\expandafter\PY@toks\fi}
\def\PY@do#1{\PY@bc{\PY@tc{\PY@ul{%
    \PY@it{\PY@bf{\PY@ff{#1}}}}}}}
\def\PY#1#2{\PY@reset\PY@toks#1+\relax+\PY@do{#2}}

\expandafter\def\csname PY@tok@gd\endcsname{\def\PY@tc##1{\textcolor[rgb]{0.63,0.00,0.00}{##1}}}
\expandafter\def\csname PY@tok@gu\endcsname{\let\PY@bf=\textbf\def\PY@tc##1{\textcolor[rgb]{0.50,0.00,0.50}{##1}}}
\expandafter\def\csname PY@tok@gt\endcsname{\def\PY@tc##1{\textcolor[rgb]{0.00,0.27,0.87}{##1}}}
\expandafter\def\csname PY@tok@gs\endcsname{\let\PY@bf=\textbf}
\expandafter\def\csname PY@tok@gr\endcsname{\def\PY@tc##1{\textcolor[rgb]{1.00,0.00,0.00}{##1}}}
\expandafter\def\csname PY@tok@cm\endcsname{\let\PY@it=\textit\def\PY@tc##1{\textcolor[rgb]{0.25,0.50,0.50}{##1}}}
\expandafter\def\csname PY@tok@vg\endcsname{\def\PY@tc##1{\textcolor[rgb]{0.10,0.09,0.49}{##1}}}
\expandafter\def\csname PY@tok@m\endcsname{\def\PY@tc##1{\textcolor[rgb]{0.40,0.40,0.40}{##1}}}
\expandafter\def\csname PY@tok@mh\endcsname{\def\PY@tc##1{\textcolor[rgb]{0.40,0.40,0.40}{##1}}}
\expandafter\def\csname PY@tok@go\endcsname{\def\PY@tc##1{\textcolor[rgb]{0.53,0.53,0.53}{##1}}}
\expandafter\def\csname PY@tok@ge\endcsname{\let\PY@it=\textit}
\expandafter\def\csname PY@tok@vc\endcsname{\def\PY@tc##1{\textcolor[rgb]{0.10,0.09,0.49}{##1}}}
\expandafter\def\csname PY@tok@il\endcsname{\def\PY@tc##1{\textcolor[rgb]{0.40,0.40,0.40}{##1}}}
\expandafter\def\csname PY@tok@cs\endcsname{\let\PY@it=\textit\def\PY@tc##1{\textcolor[rgb]{0.25,0.50,0.50}{##1}}}
\expandafter\def\csname PY@tok@cp\endcsname{\def\PY@tc##1{\textcolor[rgb]{0.74,0.48,0.00}{##1}}}
\expandafter\def\csname PY@tok@gi\endcsname{\def\PY@tc##1{\textcolor[rgb]{0.00,0.63,0.00}{##1}}}
\expandafter\def\csname PY@tok@gh\endcsname{\let\PY@bf=\textbf\def\PY@tc##1{\textcolor[rgb]{0.00,0.00,0.50}{##1}}}
\expandafter\def\csname PY@tok@ni\endcsname{\let\PY@bf=\textbf\def\PY@tc##1{\textcolor[rgb]{0.60,0.60,0.60}{##1}}}
\expandafter\def\csname PY@tok@nl\endcsname{\def\PY@tc##1{\textcolor[rgb]{0.63,0.63,0.00}{##1}}}
\expandafter\def\csname PY@tok@nn\endcsname{\let\PY@bf=\textbf\def\PY@tc##1{\textcolor[rgb]{0.00,0.00,1.00}{##1}}}
\expandafter\def\csname PY@tok@no\endcsname{\def\PY@tc##1{\textcolor[rgb]{0.53,0.00,0.00}{##1}}}
\expandafter\def\csname PY@tok@na\endcsname{\def\PY@tc##1{\textcolor[rgb]{0.49,0.56,0.16}{##1}}}
\expandafter\def\csname PY@tok@nb\endcsname{\def\PY@tc##1{\textcolor[rgb]{0.00,0.50,0.00}{##1}}}
\expandafter\def\csname PY@tok@nc\endcsname{\let\PY@bf=\textbf\def\PY@tc##1{\textcolor[rgb]{0.00,0.00,1.00}{##1}}}
\expandafter\def\csname PY@tok@nd\endcsname{\def\PY@tc##1{\textcolor[rgb]{0.67,0.13,1.00}{##1}}}
\expandafter\def\csname PY@tok@ne\endcsname{\let\PY@bf=\textbf\def\PY@tc##1{\textcolor[rgb]{0.82,0.25,0.23}{##1}}}
\expandafter\def\csname PY@tok@nf\endcsname{\def\PY@tc##1{\textcolor[rgb]{0.00,0.00,1.00}{##1}}}
\expandafter\def\csname PY@tok@si\endcsname{\let\PY@bf=\textbf\def\PY@tc##1{\textcolor[rgb]{0.73,0.40,0.53}{##1}}}
\expandafter\def\csname PY@tok@s2\endcsname{\def\PY@tc##1{\textcolor[rgb]{0.73,0.13,0.13}{##1}}}
\expandafter\def\csname PY@tok@vi\endcsname{\def\PY@tc##1{\textcolor[rgb]{0.10,0.09,0.49}{##1}}}
\expandafter\def\csname PY@tok@nt\endcsname{\let\PY@bf=\textbf\def\PY@tc##1{\textcolor[rgb]{0.00,0.50,0.00}{##1}}}
\expandafter\def\csname PY@tok@nv\endcsname{\def\PY@tc##1{\textcolor[rgb]{0.10,0.09,0.49}{##1}}}
\expandafter\def\csname PY@tok@s1\endcsname{\def\PY@tc##1{\textcolor[rgb]{0.73,0.13,0.13}{##1}}}
\expandafter\def\csname PY@tok@kd\endcsname{\let\PY@bf=\textbf\def\PY@tc##1{\textcolor[rgb]{0.00,0.50,0.00}{##1}}}
\expandafter\def\csname PY@tok@sh\endcsname{\def\PY@tc##1{\textcolor[rgb]{0.73,0.13,0.13}{##1}}}
\expandafter\def\csname PY@tok@sc\endcsname{\def\PY@tc##1{\textcolor[rgb]{0.73,0.13,0.13}{##1}}}
\expandafter\def\csname PY@tok@sx\endcsname{\def\PY@tc##1{\textcolor[rgb]{0.00,0.50,0.00}{##1}}}
\expandafter\def\csname PY@tok@bp\endcsname{\def\PY@tc##1{\textcolor[rgb]{0.00,0.50,0.00}{##1}}}
\expandafter\def\csname PY@tok@c1\endcsname{\let\PY@it=\textit\def\PY@tc##1{\textcolor[rgb]{0.25,0.50,0.50}{##1}}}
\expandafter\def\csname PY@tok@kc\endcsname{\let\PY@bf=\textbf\def\PY@tc##1{\textcolor[rgb]{0.00,0.50,0.00}{##1}}}
\expandafter\def\csname PY@tok@c\endcsname{\let\PY@it=\textit\def\PY@tc##1{\textcolor[rgb]{0.25,0.50,0.50}{##1}}}
\expandafter\def\csname PY@tok@mf\endcsname{\def\PY@tc##1{\textcolor[rgb]{0.40,0.40,0.40}{##1}}}
\expandafter\def\csname PY@tok@err\endcsname{\def\PY@bc##1{\setlength{\fboxsep}{0pt}\fcolorbox[rgb]{1.00,0.00,0.00}{1,1,1}{\strut ##1}}}
\expandafter\def\csname PY@tok@mb\endcsname{\def\PY@tc##1{\textcolor[rgb]{0.40,0.40,0.40}{##1}}}
\expandafter\def\csname PY@tok@ss\endcsname{\def\PY@tc##1{\textcolor[rgb]{0.10,0.09,0.49}{##1}}}
\expandafter\def\csname PY@tok@sr\endcsname{\def\PY@tc##1{\textcolor[rgb]{0.73,0.40,0.53}{##1}}}
\expandafter\def\csname PY@tok@mo\endcsname{\def\PY@tc##1{\textcolor[rgb]{0.40,0.40,0.40}{##1}}}
\expandafter\def\csname PY@tok@kn\endcsname{\let\PY@bf=\textbf\def\PY@tc##1{\textcolor[rgb]{0.00,0.50,0.00}{##1}}}
\expandafter\def\csname PY@tok@mi\endcsname{\def\PY@tc##1{\textcolor[rgb]{0.40,0.40,0.40}{##1}}}
\expandafter\def\csname PY@tok@gp\endcsname{\let\PY@bf=\textbf\def\PY@tc##1{\textcolor[rgb]{0.00,0.00,0.50}{##1}}}
\expandafter\def\csname PY@tok@o\endcsname{\def\PY@tc##1{\textcolor[rgb]{0.40,0.40,0.40}{##1}}}
\expandafter\def\csname PY@tok@kr\endcsname{\let\PY@bf=\textbf\def\PY@tc##1{\textcolor[rgb]{0.00,0.50,0.00}{##1}}}
\expandafter\def\csname PY@tok@s\endcsname{\def\PY@tc##1{\textcolor[rgb]{0.73,0.13,0.13}{##1}}}
\expandafter\def\csname PY@tok@kp\endcsname{\def\PY@tc##1{\textcolor[rgb]{0.00,0.50,0.00}{##1}}}
\expandafter\def\csname PY@tok@w\endcsname{\def\PY@tc##1{\textcolor[rgb]{0.73,0.73,0.73}{##1}}}
\expandafter\def\csname PY@tok@kt\endcsname{\def\PY@tc##1{\textcolor[rgb]{0.69,0.00,0.25}{##1}}}
\expandafter\def\csname PY@tok@ow\endcsname{\let\PY@bf=\textbf\def\PY@tc##1{\textcolor[rgb]{0.67,0.13,1.00}{##1}}}
\expandafter\def\csname PY@tok@sb\endcsname{\def\PY@tc##1{\textcolor[rgb]{0.73,0.13,0.13}{##1}}}
\expandafter\def\csname PY@tok@k\endcsname{\let\PY@bf=\textbf\def\PY@tc##1{\textcolor[rgb]{0.00,0.50,0.00}{##1}}}
\expandafter\def\csname PY@tok@se\endcsname{\let\PY@bf=\textbf\def\PY@tc##1{\textcolor[rgb]{0.73,0.40,0.13}{##1}}}
\expandafter\def\csname PY@tok@sd\endcsname{\let\PY@it=\textit\def\PY@tc##1{\textcolor[rgb]{0.73,0.13,0.13}{##1}}}

\def\PYZus{\char`\_}
\def\PYZob{\char`\{}
\def\PYZcb{\char`\}}

\def\PYZgt{\char`\>}
\def\PYZsh{\char`\#}

\def\PYZhy{\char`\-}
\def\PYZsq{\char`\'}
\def\PYZdq{\char`\"}

\makeatother

\hypersetup{
  pdfinfo={
    Title={The Illustris Simulation: Public Data Release},
    Author={Dylan Nelson},
    Subject={},
    Keywords={}
  }
  pdfcreator={},
  pdfproducer={}
}

\titlespacing\subsection{0pt}{12pt plus 4pt minus 2pt}{6pt plus 2pt minus 2pt}
\titlespacing\subsubsection{0pt}{12pt plus 4pt minus 2pt}{6pt plus 2pt minus 2pt}

\begin{document}

\begin{frontmatter}

\title{The Illustris Simulation: Public Data Release\tnoteref{t1}}

\author[cfa]{Dylan Nelson\corref{cor1}}
\ead{dnelson@cfa.harvard.edu}

\author[cfa]{Annalisa Pillepich}
\author[columbia,cfa]{Shy Genel\fnref{fn1}}
\author[mit]{Mark Vogelsberger}
\author[hits,heidelberg]{Volker Springel}
\author[mit,caltech]{Paul Torrey}
\author[cfa]{Vicente Rodriguez-Gomez}
\author[cambridge]{Debora Sijacki}
\author[stsci]{Gregory F. Snyder}
\author[mit]{Brendan Griffen}
\author[mit]{Federico Marinacci}
\author[maryland]{Laura Blecha\fnref{fn2}}
\author[riverside]{Laura Sales}
\author[hits]{Dandan Xu}
\author[cfa]{Lars Hernquist}

\cortext[cor1]{Corresponding author}
\fntext[fn1]{Hubble Fellow}
\fntext[fn2]{Einstein Fellow}
\tnotetext[t1]{Permanently available at \href{http://www.illustris-project.org/data/}{www.illustris-project.org/data}}

\address[cfa]{Harvard-Smithsonian Center for Astrophysics, 60 Garden Street, Cambridge, MA, 02138, USA}
\address[columbia]{Department of Astronomy, Columbia University, 550 West 120th Street, New York, NY, 10027, USA}
\address[mit]{Kavli Institute for Astrophysics and Space Research, Department of Physics, MIT, Cambridge, MA, 02139, USA}
\address[hits]{Heidelberg Institute for Theoretical Studies, Schloss-Wolfsbrunnenweg 35, 69118 Heidelberg, Germany}
\address[heidelberg]{Zentrum f\"{u}r Astronomie der Universit\"{a}t Heidelberg, ARI, M\"{o}nchhofstr. 12-14, 69120 Heidelberg, Germany}
\address[cambridge]{Institute of Astronomy and Kavli Institute for Cosmology, University of Cambridge, Madingley Road, Cambridge CB3 0HA, UK}
\address[caltech]{TAPIR, Mailcode 350-17, California Institute of Technology, Pasadena, CA 91125, USA}
\address[stsci]{Space Telescope Science Institute, 3700 San Martin Dr, Baltimore, MD 21218}
\address[riverside]{Department of Physics and Astronomy, University of California, Riverside, 900 University Avenue, Riverside, CA 92521, USA}
\address[maryland]{University of Maryland, College Park, Department of Astronomy and Joint Space Science Institute}

\begin{abstract}
We present the full public release of all data from the Illustris simulation project. 
Illustris is a suite of large volume, cosmological hydrodynamical simulations run with the 
moving-mesh code {\sc Arepo} and including a comprehensive set of physical models critical 
for following the formation and evolution of galaxies across cosmic time. Each simulates a 
volume of (106.5 Mpc)$^3$ and self-consistently evolves five different types of resolution 
elements from a starting redshift of $z=127$ to the present day, $z=0$. These components are: 
dark matter particles, gas cells, passive gas tracers, stars and stellar wind particles, and 
supermassive black holes. This data release includes the snapshots at all 136 available redshifts, 
halo and subhalo catalogs at each snapshot, and two distinct merger trees. Six primary 
realizations of the Illustris volume are released, including the flagship Illustris-1 run. 
These include three resolution levels with the fiducial ``full'' baryonic physics model, and a 
dark matter only analog for each. In addition, we provide four distinct, high time resolution, 
smaller volume ``subboxes''. The total data volume is $\sim$265 TB, including $\sim$800 full volume 
snapshots and $\sim$30,000 subbox snapshots. We describe the released data products as 
well as tools we have developed for their analysis. All data may be directly downloaded in its 
native HDF5 format. Additionally, we release a comprehensive, web-based API which allows 
programmatic access to search and data processing tasks. In both cases we provide example scripts 
and a getting-started guide in several languages: currently, IDL, Python, and Matlab. This paper 
addresses scientific issues relevant for the interpretation of the simulations, serves 
as a pointer to published and on-line documentation of the project, describes planned future 
additional data releases, and discusses technical aspects of the release.
\end{abstract}

\begin{keyword}
methods: data analysis \sep methods: numerical \sep galaxies: formation \sep galaxies: evolution \sep data management systems \sep data access methods
%distributed architectures
\end{keyword}

\end{frontmatter}

%----------------------------------------------------------------
% Introduction
%----------------------------------------------------------------

\section{Introduction}

Our theoretical understanding of the origin and evolution of cosmic structure throughout the universe is increasingly propelled forward by large, numerical simulations. From humble beginnings \citep[e.g.][]{ps74,davis85}, dark matter only N-body simulations of pure gravitational dynamics have reached a state of maturity and extreme scale \citep[e.g.][]{kim11,skillman14}. They form a foundation in our understanding of the $\Lambda$CDM cosmological model, including the nature of both dark matter and dark energy. Yet, such DM-only simulations have a fundamental limitation -- they cannot provide any direct predictions for baryonic components of the universe: gas, stars, and black holes. While dark matter halo collapse forms the back bone of structure formation, the majority of observational astronomy is based on the properties of the baryons.

The natural successor to dark matter only N-body simulations are cosmological hydrodynamical simulations \citep[e.g.][]{katz92}, which model the coupled evolution of dark matter and cosmic gas.  Hydrodynamical simulations can also account for diverse phenomena such as the formation of stars, the growth of supermassive black holes, the energetic feedback processes arising from both populations, the production and distribution of heavy elements, and so forth.  Modern efforts are now able to capture cosmological scales of $\gtrsim$100 Mpc, while simultaneously resolving the internal structure of individual galaxies at $\lesssim$1 kpc scales (Horizon-AGN: \citealt{dubois14}, MassiveBlack-II: \citealt{khandai14}, Illustris: \citealt{vog14a}, EAGLE: \citealt{schaye15}).  These simulations yield verifiable predictions or models for a wide range of interesting astrophysical problems including the spin alignment of galaxies on large scales \citep[e.g.][]{hahn10}, the distribution of neutral hydrogen \citep[e.g.][]{bird14,rahmati15}, or the impact of baryons on the structure of dark matter haloes \citep[e.g.][]{schaller14a}.

Observational data focused on the large-scale structure of the universe and the properties of galaxies across cosmic time also continue to increase. Surveys such as SDSS \citep{york00}, DEEP2 \citep{davis03}, CANDELS \citep{grogin11}, and 3D-HST \citep{brammer12} provide local and high redshift  measurements of the statistical properties of galaxy populations. Future instruments such as LSST \citep{lsst09} and surveys such as DES \citep{des05} will provide increasingly precise observational constraints for theoretical models.

To confront theory and observation, the public dissemination of data from both sides is crucial. Efforts based on the availability of ubiquitous international networks began with the highly successful SDSS SkyServer \citep{szalay00,szalay02a}, which addressed the problems of how remote users could mine data from large datasets \citep{gray02,szalay02b}. The approach, which continues to this day, is based on user written SQL queries executed against a large relational database system -- query responses can be thought of as both search results and data extraction. Simple queries with near-instantaneous return, as well as long, queued job queries with results saved into temporary storage are supported.

The Millennium simulation \citep{spr05c} public data release was the first large effort from the theoretical side. Modeled on the SDSS approach, the primary data products were stored in a relational database, which users could search and extract data from using raw SQL queries \citep{lemson06}. The focus is on the halo and subhalo catalogs, their merger trees, and various post-processed galaxy property catalogs computed with semi-analytical models.
It has been continually extended with additional simulations, data products, and capabilities. The Millennium-II simulation \citep{boylan09,guo11} was included, and the idea of the ``virtual observatory'' (VO) was realized with \cite{overzier13}. These efforts have occasionally implemented ideas for incorporating theory within the existing VO framework \citep{lemson09,lemson14}. More generally, the Theoretical Astrophysical Observatory \citep[TAO][]{bernyk14} was also targeted at providing mock observations of simulated galaxy and galaxy survey data.

Other dark matter only simulations have adopted similar approaches. The Bolshoi and MultiDark simulations \citep{klypin11} were released under a common database \citep{riebe13}, now called CosmoSim. The Dark Energy Universe Simulation \citep[DEUS][]{rasera10} data is available online, as are some data from the MICE simulations \citep{crocce10} through the CosmoHub database. In contrast, the MassiveBlack-II (hydrodynamical) simulation \citep{khandai14} made group catalogs available for direct download. Most recently, the Dark Sky simulation has likewise avoided the database and SQL query framework in favor of direct web access to binary data \citep{skillman14}.

In releasing the Illustris simulation data, we adopt a similar approach, offering direct online access to all snapshot, group catalog, merger tree, and supplementary data catalog files. In addition, we develop a web-based API which allows users to perform many common tasks without the need to download any full data files. These include searching over the group catalogs, extracting particle data from the snapshots, accessing individual merger trees, and requesting visualization and further data analysis functions. Extensive documentation and programmatic examples (in IDL, Python, and Matlab) are provided.

This paper is intended primarily as a guide for users of the Illustris simulation data. In Section \ref{sSims} we give an overview of the simulations. Section \ref{sDataProducts} describes the data products, and Section \ref{sDataAccess} discusses methods for data access. Section \ref{sImplementation} describes technical details related to the architecture and implementation of the data release itself. In Section \ref{sRemarks} we present some scientific remarks and cautions for Illustris, while in Section \ref{sCommunity} we discuss community considerations including citation. In Section \ref{sConclusions} we summarize. Appendices A through C provide descriptions of all relevant data fields, while Appendix D presents several code examples for the API.

%----------------------------------------------------------------
% Description of the Simulations
%----------------------------------------------------------------

\section{Description of the Simulations} \label{sSims}

\begin{table*}[htb!]
\small
  \caption{The most important numerical parameters for the six full volume runs. Gravitational softenings for all particle types other than DM are comoving kpc (with value equal to that of the DM) until $z=1$ after which they are fixed to their $z=1$ values, such that at $z=0$ they have half the softening length as the DM.
$ m_{\rm baryon} $ is the ``target gas mass'' (i.e. only the mean mass).
The number of gas cells equals the $N_{\rm GAS}$ value only in the initial conditions, the number will then drop as stars and black holes form. Moreover, the total number of baryonic particles (gas cells + star particles + wind particles + black holes) is also not conserved since gas cells can be refined/de-refined to keep their mass within a factor of 2 around $m_{\rm baryon}$. In contrast, the total number of tracers and dark matter particles are both conserved for the duration of the simulation.}
  \label{table_sims}
  \begin{center}
\renewcommand{\arraystretch}{1.5}
    \begin{tabular}{llcccccccccc}
     \hline
 Run Name &
 Alt. Name &
 Volume  &
 $L_{\rm box}$ &
 $ N_{\rm GAS} $ &
 $ N_{\rm TR} $ &
 $ N_{\rm DM} $ &
 $ \epsilon_{ \rm baryon}$ &
 $ \epsilon_{DM}$ &
 $ m_{baryon} $ &
 $ m_{DM} $  \\ 
 &
 &
 $ [ {\rm Mpc}^3 ] $ &
 $ [ {\rm Mpc}/h ] $ &
 &
 &
 &
 $ [{\rm kpc}] $ &
 $ [{\rm kpc}] $ &
 $[ {\rm M}_\odot ]$ &
 $[ {\rm M}_\odot ] $ \\ \hline\hline
Illustris-1      & L75n1820FP  & 106.5$^3$ & 75 & 1820$^3$ & 1820$^3$ & 1820$^3$ & 0.7 & 1.4 & $1.6 \times 10^6$ & $6.3 \times 10^6$ \\
Illustris-2      & L75n910FP   & 106.5$^3$ & 75 & 910$^3$  &  910$^3$ &  910$^3$ & 1.4 & 2.8 & $1.0 \times 10^7$ & $5.0 \times 10^7$ \\
Illustris-3      & L75n455FP   & 106.5$^3$ & 75 & 455$^3$  &  455$^3$ &  455$^3$ & 2.8 & 5.7 & $8.0 \times 10^8$ & $4.0 \times 10^8$ \\
Illustris-1-Dark & L75n1820DM  & 106.5$^3$ & 75 & 0 & 0    & 1820$^3$ &        - & 1.4 &   - & $7.6 \times 10^6$ \\
Illustris-2-Dark & L75n910DM   & 106.5$^3$ & 75 & 0 & 0    &  910$^3$ &        - & 2.8 &   - & $6.0 \times 10^7$ \\
Illustris-3-Dark & L75n455DM   & 106.5$^3$ & 75 & 0 & 0    &  455$^3$ &        - & 5.7 &   - & $4.8 \times 10^8$ \\ 
\hline
    \end{tabular}
  \end{center}
\end{table*}

The Illustris Project is a series of hydrodynamical simulations of a (106.5 Mpc)$^3$ cosmological volume that follow the evolution of dark matter, cosmic gas, stars, and super massive black holes from a starting redshift of $z=127$ to the present day, $z=0$. It includes three runs at increasing resolution levels, Illustris-(1,2,3), where Illustris-1 is the flagship, highest-resolution box. Each has been simulated including a fiducial ``full'' baryonic physics model, as well as a dark-matter only analog, Illustris-(1,2,3)-Dark. \cite{vog14a,vog14b,genel14,sijacki14} have presented the Illustris simulations and their galaxy and black hole populations, both at $z=0$ as well as at higher redshifts. In what follows, we summarize the most relevant features.

\begin{figure*}[htb!]
\centerline{\includegraphics[angle=0,width=7.3in]{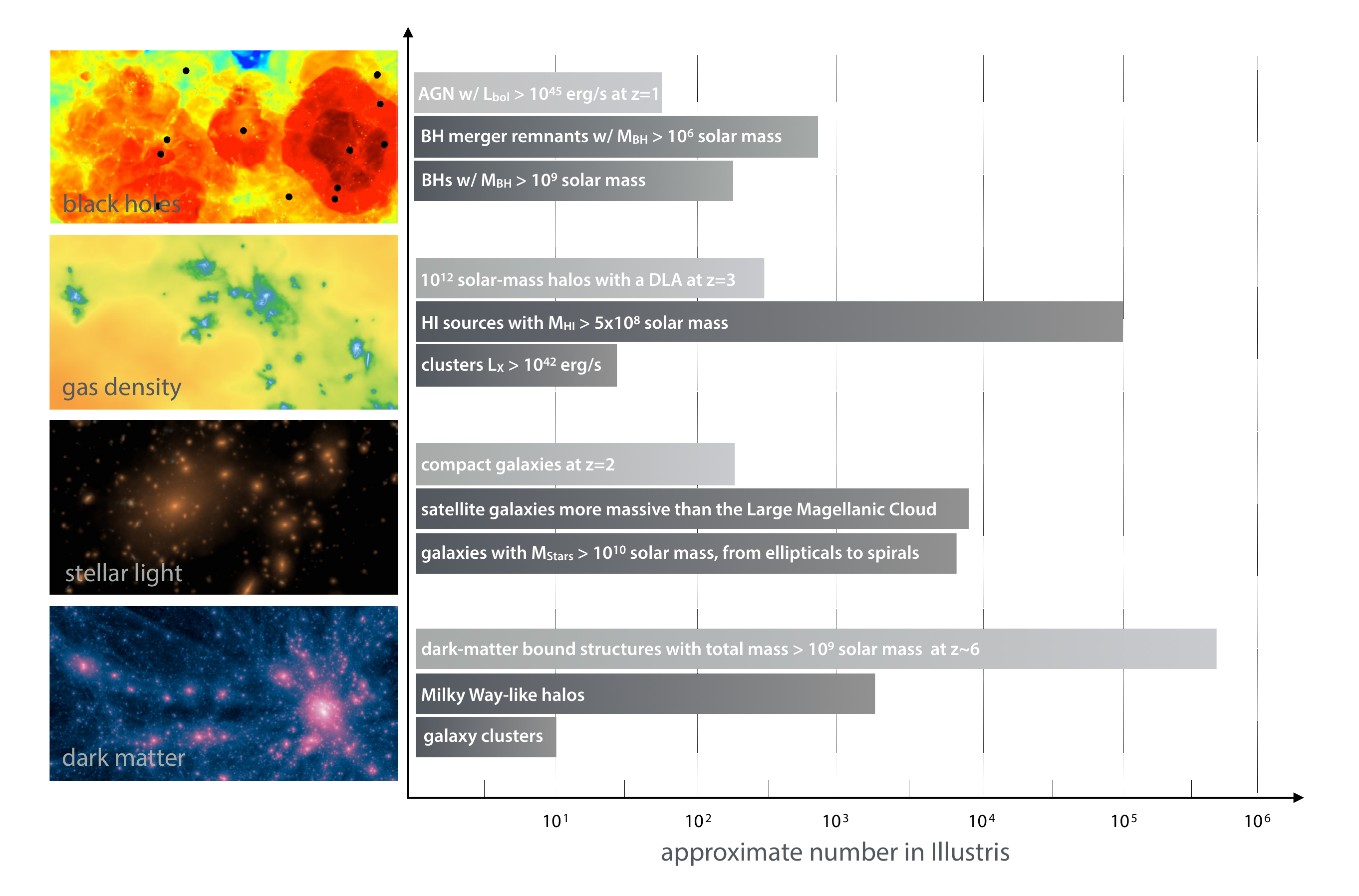}}
\caption{ Overview of the variety of galaxy formation and evolution phenomena accessible in the Illustris simulations. A few classes of interesting objects are listed for each of the four mass components present in the simulation: dark matter, stars, gas, and black holes. These are visualized on the left column, for different volumes and spatial scales, as dark-matter density, stellar light, gas density and gas temperature maps, with black holes denoted as black dots. The approximate number present in the Illustris-1 volume is given (from bottom to top), for a) galaxy clusters at $z=0$ with total mass $M_{200c}> 10^{14} {\rm M}_\odot$; b) Milky Way-like halos at $z=0$ ($ 6 \times 10^{11} < M_{200c}< 2 \times 10^{12} {\rm M}_\odot$); c) gravitationally-bound objects (dark or luminous) resolved with more than a thousand particles at the end of the reionization epoch; d) galaxies at $z=0$ with stellar mass exceeding $10^{10} {\rm M}_\odot$, including both centrals and satellites, from elliptical to disk morphologies; e) satellite galaxies at $z=0$ more massive than the Large Magellanic Cloud (stellar mass $> 1.5 \times 10^9 {\rm M}_\odot$), in any mass host; f) massive, compact galaxies at $z=2$ according to the selection of \cite{barro13}; g) clusters of galaxies at $z=0$ emitting in the X-rays with luminosity exceeding $10^{42}$ erg/s; h) sources at $z=0$ with neutral hydrogen mass exceeding $5 \times 10^8 {\rm M}_\odot$; i) $10^{12} {\rm M}_\odot$ halos at $z=3$ with at least a damped Lyman-alpha system (HI column density $> 10^{20.3} {\rm cm}^{-2}$) within $50 {\rm kpc}$; j) black holes at $z=0$ more massive than $10^9 {\rm M}_\odot$; k) black-hole merger remnants at $z=0$ , i.e. sub grid black-hole binaries with $M_{\rm BH} > 10^6 {\rm M}_\odot$ for each BH and 1 Gyr delay between the simulation BH merger time and the actual BH merger; l) AGNs at $z=1$ with bolometric luminosity greater than $10^{45}$ erg/s.
 \label{fig_objects}} 
\end{figure*}

In Table \ref{table_sims} we provide an overview of the specifications of the six Illustris runs, including the computational volume, gravitational softening lengths, and masses of the different particle/cell types, which collectively indicate the resolution and dynamic range achieved.
To emphasize the variety of galaxy formation and evolution phenomena which can be addressed with the Illustris simulations, in Figure \ref{fig_objects} we give the approximate number of a selection of interesting astrophysical objects that can be found in the simulated box, from dark-matter dominated halos at $z=0$ to luminous active galactic nuclei (AGN) at higher redshifts. 

A series of analyses based on the Illustris suite have already been performed. These include 1) comparisons to observations and studies of the impact of different feedback models on the distribution and content of gas on large scales, within halos and in the circumgalactic regime \citep{bird14,bird15,nelson15a,suresh15,bogdan15}; 2) characterizations of the properties of galactic stellar halos \citep{pillepich14}, of the satellite populations across host masses \citep{sales15}, of the star formation histories \citep{sparre15} and of the morphologies and angular-momentum build up of Illustris galaxies \citep{torrey15,snyder15,genel15}; 3) applications of shock finder algorithms \citep{schaal15}; 4) analyses on the formation of massive, compact galaxies at high redshifts \citep{wellons15}; 5) quantification of the galaxy merger rates \citep{rodrig15}, and 6) applications of post-processing radiative transfer algorithms in the study of cosmic reionization \citep{bauer15}.

\subsection{Physical Models and Numerical Methods}

All of the ``full physics'' Illustris runs contain the following physical components:
(1) Primordial and metal-line radiative cooling in the presence of a redshift-dependent, spatially uniform, ionizing UV background field, with self-shielding corrections.
(2) Stochastic star formation in dense gas.
(3) Pressurization of the ISM due to unresolved supernovae using an effective equation of state model of a two-phase medium.
(4) Stellar evolution with the associated mass loss (gas recycling) and chemical enrichment, taking into account SN Ia/II and AGB stars.
(5) Galactic-scale outflows with an energy-driven, kinetic wind scheme.
(6) Seeding and growth of supermassive black holes.
(7) Feedback from AGN in both quasar and radio (bubble) modes, as well as modifications to the cooling curve of nearby gas due to radiation proximity effects.
For complete details on the behavior, implementation, parameter selection, and validation of these physical models, see 
\cite{vog13}, which describes the feedback models, and \cite{torrey14}, which compares the model output with observations from $z=0$ to $z=3$.

The Illustris simulations employ the {\sc Arepo} code \linebreak\citep{spr10} which evolves the equations of continuum hydrodynamics coupled with self-gravity. The spatial discretization of the fluid is 
provided by an unstructured, moving, Voronoi tessellation. On the volumes defined by individual cells Godunov's method 
is employed, with a directionally unsplit MUSCL-Hancock scheme and an exact Riemann solver. The Voronoi mesh is generated from 
a set of control points which move with the local fluid velocity modulo mesh regularization corrections. Gravitational forces 
are computed using the Tree-PM approach, with long-range forces calculated with a Fourier particle-mesh method, and short-range 
forces with a hierarchical tree algorithm. The code is second order in space, and with hierarchical adaptive time-stepping, 
also second order in time. During the simulation we employ the Monte Carlo tracer particle scheme \citep{genel13} to 
follow the Lagrangian evolution of baryons.

\begin{figure*}[htb!]
\centerline{\includegraphics[angle=0,width=7.3in]{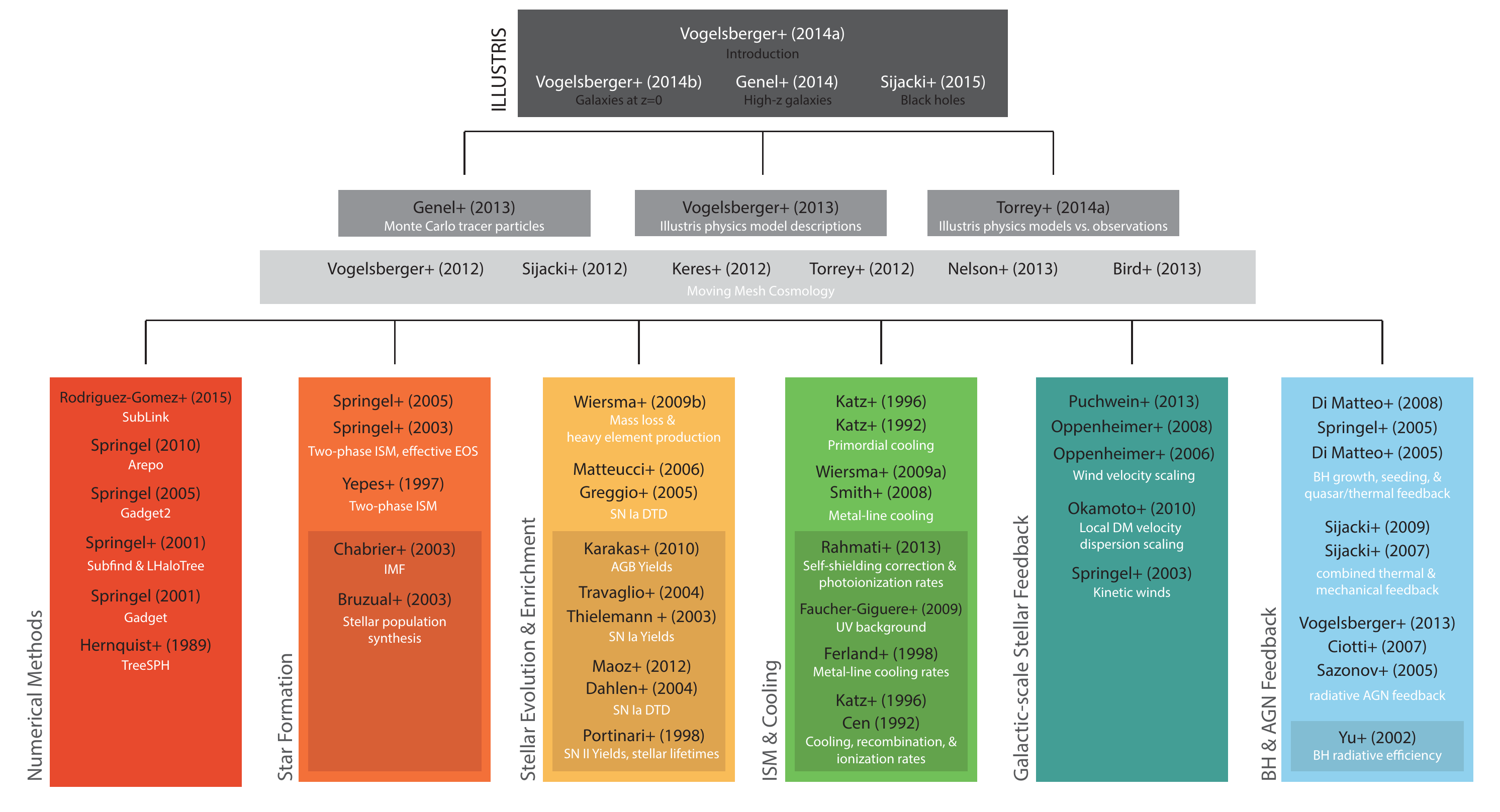}}
\caption{ Reference tree for the major components of Illustris, including both numerical methods and physical models. 
Each paper links to its arXiv or ADS entry. We generally include both models and methods which were directly implemented in Illustris, 
while entries in the dark subboxes indicate model data inputs. The references are, for the second row: \cite{genel13,vog13,torrey14}.
The moving mesh cosmology series: \cite{vog12,sijacki12,keres12,torrey12,nelson13,bird13}.
Numerical methods: \cite{rodrig15,spr10,spr05b,spr01,spr01b,hernquist89}.
Star formation: \cite{spr05,spr03,yepes97,chabrier03,bruzual03}. 
Stellar evolution and enrichment: \cite{wiersma09b,matteucci06,greggio05,karakas10,travaglio04,thiel03,maoz12,dahlen04,portinari98}. 
ISM and cooling: \cite{katz92,katz96,wiersma09,smith08,rahmati13,fg09,ferland98,katz96,cen92}.
Galactic-scale stellar feedback: \cite{puchwein13,opp08,opp06,okamoto10,spr03}.
BH and AGN feedback: \cite{dimatteo08,spr05,dimatteo05,sijacki07,sijacki09,vog13,ciotti07,sazonov05,yu02}.
 \label{fig_references}} 
\end{figure*}

In terms of both physical models and numerical methods, the Illustris simulations rely on a substantial foundation of previous work.
In Figure \ref{fig_references}, we provide an abridged reference tree covering both the physical models and numerical methods. 
The papers along any given branch are essential for understanding the details and limitations of the data released here.

%----------------------------------------------------------------
% Data Products
%----------------------------------------------------------------

\section{Data Products} \label{sDataProducts}

In this data release we give public access to all 136 snapshots between redshift $z=40$ and redshift zero of the Illustris cosmological volume.
This is a periodic box of 106.5 Mpc per side, including up to five types of resolution elements (dark matter particles, gas cells, gas tracers, stellar and stellar wind particles, and black hole sinks).
The same volume is available at high (Illustris-1), intermediate (Illustris-2), and low (Illustris-3) resolution.
For each resolution, realizations exist with our fiducial, full physics models (``Illustris''), as well as dark matter only analogs (``Illustris Dark'').
For all six runs, at every snapshot, two types of group catalogs are provided: friends-of-friends (FoF) halo catalogs, and {\sc Subfind} subhalo catalogs.
In postprocessing, these catalogs are used to generate two distinct merger trees, which are both released: {\sc SubLink}, and {\sc LHaloTree}.
Finally, supplementary data catalogs are released for selected snapshots and runs.
At present, these are focused on the stellar properties of Illustris-1 galaxies at $z=0$, and include mock multi-band images, photometric non-parametric morphological estimates, circularities, angular momenta, and axis ratio measurements.
All these data types are described below (snapshots, group catalogs, merger trees, and supplementary catalogs).
In the near future we plan to release {\sc Rockstar} group catalogs and the associated {\sc Consistent-Trees} merger histories, together with expanded and new supplementary catalogs, with corresponding documentation.

\subsection{Snapshots}

\subsubsection{Snapshot Organization}

There are 136 snapshots stored for every run. These include all particles/cells in the whole volume. 
The full snapshot listings, spacings and redshifts can be found online. A partial listing is provided in Table \ref{table_snaps}.
Every snapshot is stored in a series of ``chunks'', i.e. more manageable, smaller-size files. The number of chunks per snapshots is different for the different runs, and is given in Table \ref{table_chunks}.

\begin{table}[htb!]
\footnotesize
  \caption{Abridged snapshot list for all six runs. The output times correspond to the set of 128 output redshifts 
  used by the Aquarius project \citep{spr08}, augmented by 8 additional saves at integer redshifts.}
  \label{table_snaps}
  \begin{center}
\renewcommand{\arraystretch}{1.3}
    \begin{tabular}{cll}
    \hline
Snapshot & Scale factor & Redshift \\ \hline\hline
0  & 0.020932 & 46.773 \\
32 & 0.090937 & 9.9966 \\
45 & 0.14264 & 6.0108 \\
54 & 0.19968 & 4.0079 \\
60 & 0.24949 & 3.0081 \\
68 & 0.33311 & 2.002 \\
85 & 0.50068 & 0.9973 \\
135 & 1.0 & 0.0 \\
 \hline
    \end{tabular}
  \end{center}
\end{table}

\begin{figure}[tb!]
\centerline{\includegraphics[angle=0,width=3.4in]{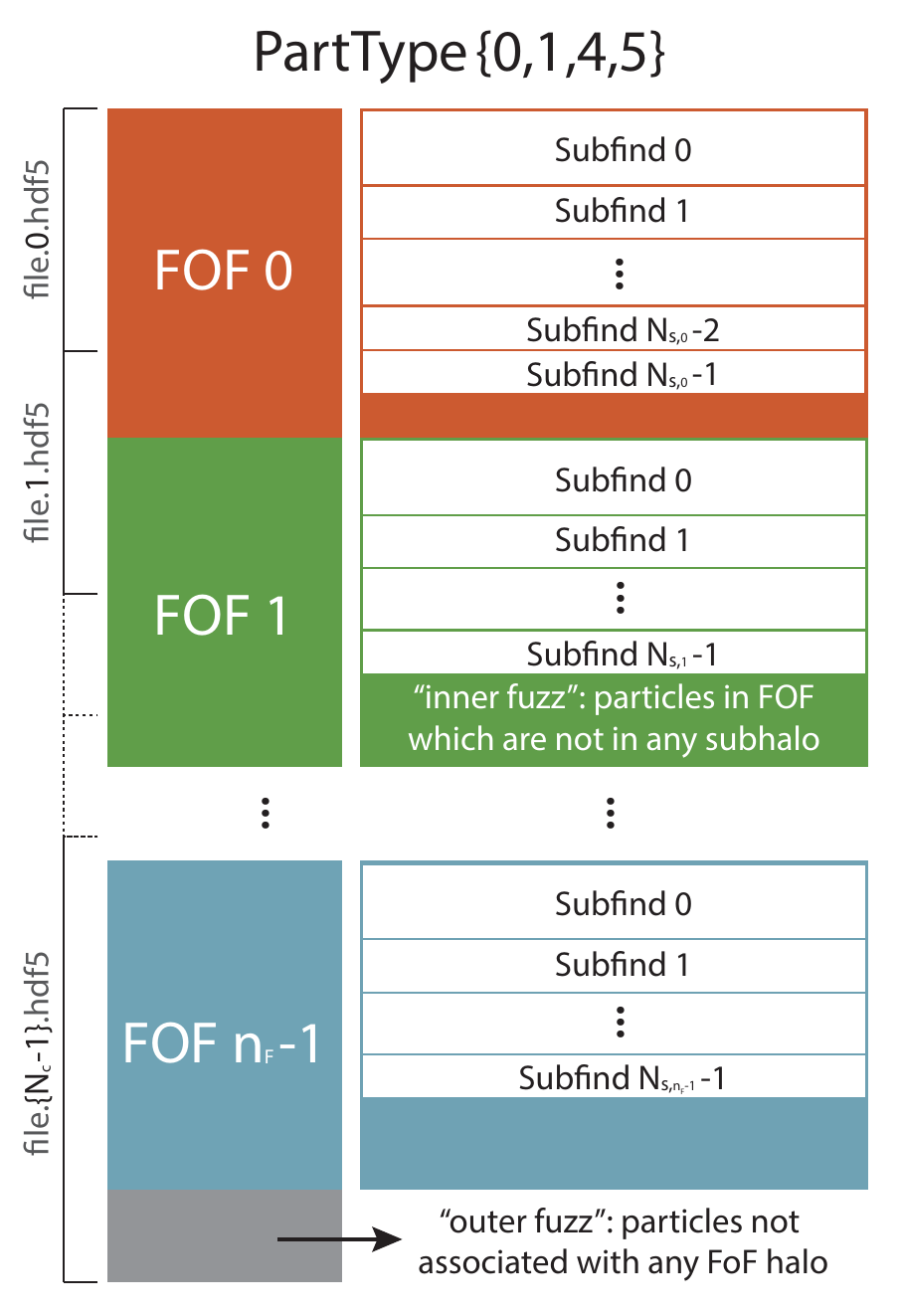}}
\caption{ Schematic diagram of the organization of particle/cell data within the snapshots for a single particle type. Within a type, particle order is determined by a global sort of the following fields in this order: FoF group number, {\sc Subfind} subhalo number, binding energy, nearest FoF group number. This implies that FOF halos are contiguous, although they can span file chunks. {\sc Subfind} subhalos are only contiguous within a single group, being separated between groups by an ``inner fuzz'' of all FOF particles not bound to any subhalo. Here $N_c$ indicates the number of file chunks, $n_F$ the number of FOF groups, and $N_{S,j}$ the number of subhalos in $j^{\rm th}$ FoF group.
 \label{fig_snap_schematic}} 
\end{figure}

The snapshot data is \textbf{not} organized according to spatial position. Rather, particles within the snapshot files are sorted according to their group/subgroup memberships, according to the FoF or {\sc Subfind} algorithms. Within each particle type, the sort order is: GroupNumber, SubgroupNumber, BindingEnergy, where particles belonging to the group but not to any of its subgroups (``fuzz'') are included after the last subgroup.
Figure \ref{fig_snap_schematic} provides a schematic view of the particle organization within a snapshot, for \textit{one particle type}. The truncation of a snapshot in chunks is arbitrary, thus halos may happen to be stored across multiple, subsequent chunks. Similarly, the different particle types of a halo can be stored in different sets of chunks.

\subsubsection{Snapshot Contents}

Every HDF5 snapshot contains a ``Header'' and 5 additional ``PartTypeX'' groups, for the following particle types (the DM only runs have a single PartType1 group):\par\nobreak

\begin{itemize}[leftmargin=2em]
\setlength\itemsep{-0.2em}
\item PartType0 - GAS
\item PartType1 - DM
\item PartType2 - (unused)
\item PartType3 - TRACERS
\item PartType4 - STARS \& WIND PARTICLES
\item PartType5 - BLACK HOLES
\end{itemize}

The most important fields of the header are given in Table \ref{table_snap_header}.
The complete snapshot field listings, including dimensions, units and descriptions, are given for gas in Table \ref{table_gas}, dark matter in \ref{table_dm}, tracers in \ref{table_tracers}, stars in \ref{table_stars}, and black holes in \ref{table_bhs}.

The general unit system is ${\rm kpc}/h$ for lengths, $10^{10} {\rm M}_\odot/h$ for masses, ${\rm km/s}$ for velocities.
The frequently occurring $(10^{10} {\rm M}_\odot/h) / (0.978 {\rm Gyr}/h)$ represents mass-over-time in this unit system, and multiplying by 10.22 converts to ${\rm M}_\odot/{\rm yr}$.
Comoving quantities can be converted in the corresponding physical ones by multiplying for the appropriate power of the scale factor a.
For instance, to convert a length in physical units it is sufficient to multiply it by $a$, volumes need a factor $a^3$, densities $a^{-3}$ and so on.
Note that at redshift $z=0$ the scale factor is $a=1$, so that the numerical values of comoving quantities are the same as their physical counterparts.

\subsubsection{Tracer Quantities}

Each Monte Carlo tracer particle stores 13 auxiliary values. These are updated every timestep where the tracer parent is active. Many are reset to zero immediately after they are written out to a snapshot, such that their recording duration is precisely the time interval between two successive snapshots. Some are only relevant when the tracer resides within a parent of a specific particle type (e.g. gas or star). Table \ref{table_tracer_quantities} describes these fields. As the simulations evolve, tracers are exchanged (and can therefore change their parents) in the following ways:

\begin{itemize}[leftmargin=2em]
\setlength\itemsep{0em}
\item Gas -$>$ Gas (finite volume fluxes, refinement, derefinement)
\item Gas -$>$ Stars (star formation, both spawning new stars and converting cells into stars)
\item Stars -$>$ Gas (stellar mass return)
\item Gas -$>$ Wind (galactic scale stellar winds)
\item Wind -$>$ Gas (recoupling stellar wind)
\item Gas -$>$ BHs (black hole accretion)
\item BHs -$>$ BHs (black hole mergers)
\end{itemize}

\subsubsection{Subboxes}

Four separate ``subbox'' cutouts exist, for each full physics run. These are spatial cutouts of fixed comoving size and fixed comoving coordinates. They are output at each highest timestep, that is, their time resolution is significantly better than that of the main snapshots -- see Table \ref{table_subbox1}. This can be particularly useful for certain types of analysis or particular science questions, or for time evolving visualizations. We point out two notes of caution: first, the time spacing of the subboxes is not uniform in scale factor or redshift, but scales with the time integration hierarchy of the simulation, and is thus variable, with some discrete factor of two jumps at several points during the simulations. Second, the subboxes, unlike the full box, are not periodic.

The four subboxes sample four different areas of the large box, roughly described by the environment column in Table \ref{table_subbox2}.
The particle fields are all identical to the main snapshots. However, the ordering differs. In particular, particles/cells in the subboxes are not ordered according to their group membership, as no group catalogs are available for these cutouts.

\begin{table}[tb!]
\footnotesize
  \caption{Details of the subbox snapshots. For each resolution level, from lowest to highest, the total number of subbox snapshots saved $N_{\rm snap}$. Each of the four subboxes has the same number of snapshots. The number of file pieces per snapshot $N_c$, and the approximate time resolution $\Delta t$ at three redshifts: $z=6$, $z=2$, and $z=0$.}
  \label{table_subbox1}
  \begin{center}
\renewcommand{\arraystretch}{1.5}
    \begin{tabular}{cccccc}
    \hline
Run & $N_{\rm snap}$ & $N_c$ & $\Delta t_{(z=6)}$ & $\Delta t_{(z=2)}$ & $\Delta t_{(z=0)}$ \\ \hline\hline
Illustris-3 &  1426 &  1   &  $\sim$7 Myr &  $\sim$12 Myr &  $\sim$33 Myr \\
Illustris-2 &  2265 &  16  &  $\sim$4 Myr &  $\sim$6 Myr  &  $\sim$17 Myr \\
Illustris-1 &  3976 &  512 &  $\sim$2 Myr &  $\sim$3 Myr  &  $\sim$8 Myr \\ \hline
    \end{tabular}
  \end{center}
\end{table}

\subsection{Group Catalogs}

There is one group catalog associated with each snapshot, which includes both FoF and {\sc Subfind} objects. The group files are split into a small number of sub-files, just as with the raw snapshots. Every group catalog file contains the following HDF5 groups: Header, Group, Subhalo, Offsets.
The IDs of the members of each group/subgroup are not stored in the group catalog files. Rather, particles/cells in the snapshot files are ordered according to group membership. Each group contains its total length, allowing IDs and all other fields of member particles/cells to be accessed using an offset table type approach.  This applies to subhalos as well, e.g. the subhalos belonging to group 0 are listed first.

In order to reduce confusion, we adopt the following terminology when referring to different types of objects.
``Group'', ``FoF Group'', and ``FoF Halo'' all refer to halos. ``Subgroup'', ``Subhalo'', and ``Subfind Group'' all refer to subhalos.
The first (most massive) subgroup of each halo is the ``Primary Subgroup'' or ``Central Subgroup''.
All other following subgroups within the same halo are ``Secondary Subgroups'', or ``Satellite Subgroups''.

\textbf{FoF Groups.} The Group fields are derived with a standard friends-of-friends (FoF) algorithm with linking length $b=0.2$. The FoF algorithm is run on the dark matter particles, and the other types (gas, stars, BHs) are attached to the same groups as their nearest DM particle. The fields for the FoF halo catalog are described in Table \ref{table_gc_fof}.

\textbf{Subfind Groups.} The Subhalo fields are derived with the {\sc Subfind} algorithm, last described in \cite{spr05}. In identifying gravitationally bound substructures the method considers all particle types and assigns them to subhalos as appropriate. It has undergone many modifications to add additional properties to each subhalo entry. Descriptions of all fields in this subhalo catalog are split across Tables \ref{table_gc_subfind1} and \ref{table_gc_subfind2}.

\textbf{Header and Offsets.} Table \ref{table_gc_header} describes the fields in the Header group, while Table \ref{table_gc_offsets} describes the fields in the Offsets group.
Note that we simply store the offsets here, which relate to all types of data files and not solely to the group catalogs.

\subsection{Merger Trees}

\begin{figure*}[htb!]
\centerline{\includegraphics[angle=0,width=7.3in]{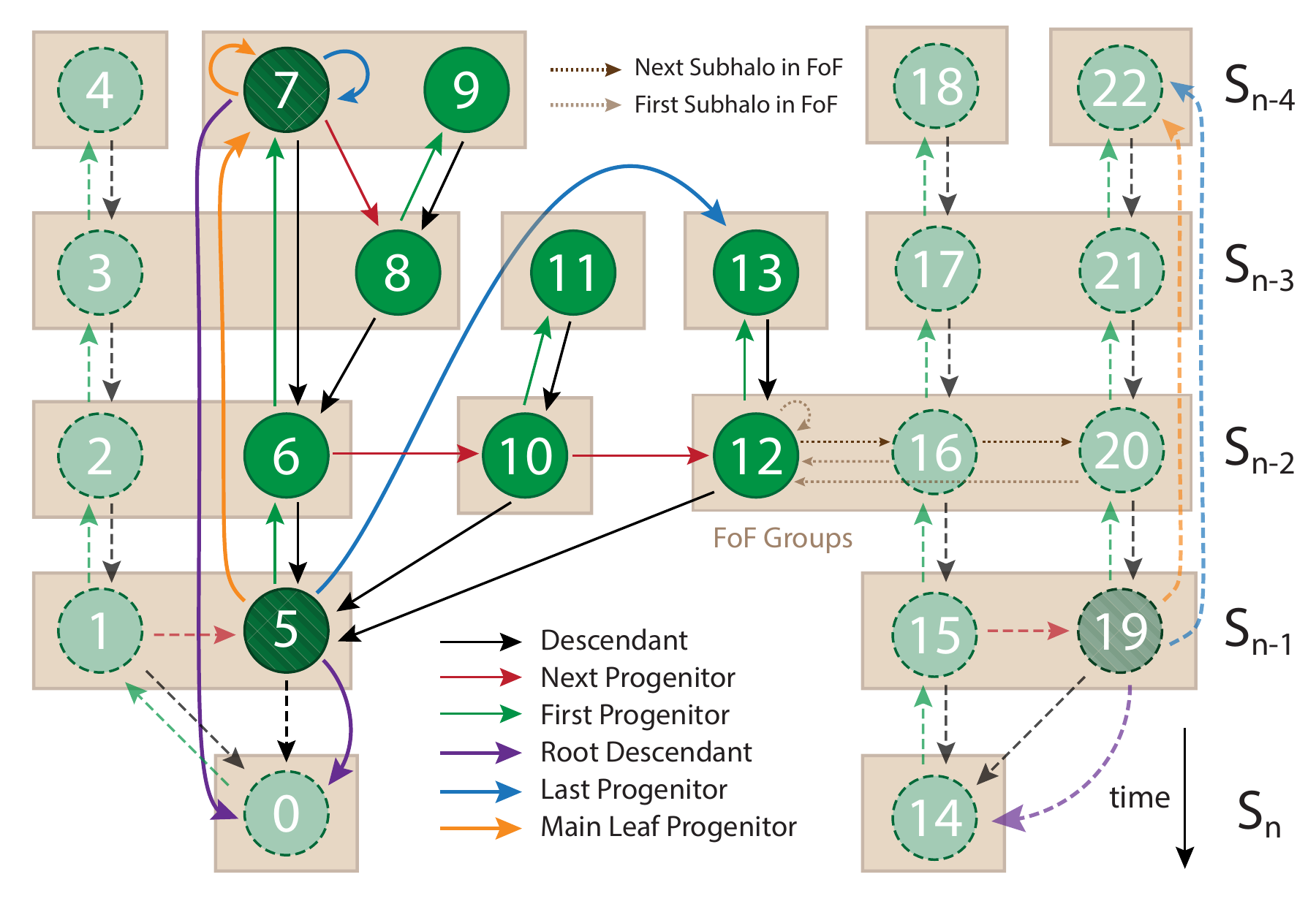}}
\caption{ Schematic diagram of the merger tree structure for both {\sc SubLink} and {\sc LHaloTree}. Both algorithms connect subhalos (i.e., {\sc Subfind} halos) across different snapshots in the simulation. Rows indicate discrete snapshots, with time increasing downwards towards redshift zero (the horizontal axis is arbitrary). Green circles represent subhalos (the nodes of the merger tree), while beige boxes indicate the grouping of the subhalos into their parent FoF groups. The most important links are for the descendant (black), first progenitor (green), and next progenitor (red), which are shown for all subhalos. The root descendant (purple), last progenitor (blue), and main leaf progenitor (orange) links exist only for the {\sc SubLink} trees, and for simplicity these last three link types are shown only for subhalos 5, 7, and 19 (darker striped circles). For exact definitions of each link type, see the corresponding tables. For more information about this figure, consult the text.
 \label{fig_tree_schematic}} 
\end{figure*}

Merger trees have been created for the various Illustris simulations using {\sc SubLink} \citep{rodrig15}, {\sc LHaloTree} \citep{spr05}, and \linebreak{\sc Consistent-Trees} (using {\sc Rockstar}, \citealt{behroozi13}, not discussed in detail here). These codes are all included in the Sussing Merger Trees comparison project \citep{srisawat13}. In the population average sense the different merger trees give similar results. In more detail, the exact merger history or mass assembly history for any given halo may differ. For a particular science goal, one type of tree may be more or less useful, and users are free to use whichever they prefer. The explicit differences between the otherwise similar {\sc LHaloTree} and {\sc SubLink} algorithms are noted below, here we detail their common features.

Figure \ref{fig_tree_schematic} shows a schematic of the structure of both the {\sc SubLink} and {\sc LHaloTree} merger trees. It is not necessary to understand the complete details of the trees to practically use them. In particular, the only critical links are the `descendant' (black), `first progenitor' (green), and `next progenitor' (red) associations. These are shown for all tree nodes in the diagram. For their exact definitions, see Tables \ref{table_sublink} and \ref{table_lhalotree}, the {\sc LHaloTree} and {\sc SubLink} tables. Walking back in time following along the main (most massive) progenitor branch consists of following the first progenitor links until they end (value equals -1). Similarly, walking forward in time along the descendants branch consists of following the descendant links until they end (value equals -1), which typically occurs at $z=0$. The full progenitor history, and not just the main branch, requires following both the first and next progenitor links. In this way the user can identify all subhalos at a previous snapshot which have a common descendant. Examples of walking the tree are provided in the example scripts.

The number inside each circle from the figure is the unique ID (within the whole simulation) of the corresponding subhalo, which is assigned in a depth-first fashion. Numbering also indicates the on-disk storage ordering for the {\sc SubLink} trees, which adopt the approach of \cite{lemson06,lemson06b}. For example, the main progenitor branch (from 5-7 in the example) and the full progenitor tree (from 5-13 in the example) are both contiguous subsets of each merger tree field, whose location and size can be calculated using these links. The ordering within a single tree in the {\sc LHaloTree} is not guaranteed to follow this scheme.

The `root descendant' (purple), `last progenitor' (blue), and `main leaf progenitor' (orange) links exist only for the {\sc SubLink} trees. For simplicity, these last three link types are shown only for nodes 5, 7, and 19 (darker striped circles). Using these links is optional, but allows efficient extraction of main progenitor branches, subtrees (i.e., the set containing a subhalo and ``all'' its progenitors), ``forward'' descendant branches, and other subsets of the tree. For their full definitions, see Table \ref{table_sublink} with the {\sc SubLink} details. 

Each subhalo spans a ``subtree'' consisting of the subhalo itself and all its progenitors. As an example, the subhalos belonging to the subtree of subhalo 5 are shown in darker green in the figure. Other subhalos not belonging to this subtree are shown in lighter green, and their links are indicated with dashed arrows. In the {\sc SubLink} trees, the subtree of any subhalo can be extracted easily using the `last progenitor' pointer. As shown in the figure, since subhalo 13 is the `last progenitor' of subhalo 5, the subtree of subhalo 5 consists of all subhalos with IDs between 5 and 13. Similarly, the main progenitor branch of any subhalo can be retrieved efficiently using the `main leaf progenitor' link.

Both {\sc SubLink} and {\sc LHaloTree} contain the links `first subhalo in FoF group' (light brown dotted arrow) and `next subhalo in FoF group' (dark brown dotted arrow), which connect subhalos that belong to the same FoF group. The FoF groups do not play a direct role in the construction of the merger tree. However, subhalos that belong to the same FoF group are also considered to be part of the same tree. As a result, two otherwise independent trees (based on the progenitor and descendant links) are considered to be the same tree if they are ``connected'' by a FoF group. This is exemplified in the figure by the FoF group containing subhalos 12, 16, and 20. This FoF group acts as a ``bridge'' between the left and right trees.

Between the otherwise similar {\sc LHaloTree} and {\sc SubLink} algorithms there are three explicit differences, in (i) the merit function used to rank descendants, (ii) the method for skipping snapshots, and (iii) the definition of the main progenitor. In both cases, descendant candidates are identified for each subhalo as those subhalos in the following snapshot(s) that have common particles with the subhalo in question. These candidates are given a score based on a merit function which takes into account the binding energy rank of each matched particle. In this way, preference is given to tracking the fate of the inner parts of a structure, which may survive for a long time upon infall into a bigger halo, even though much of the mass in the outer parts can be quickly stripped. The unique descendant of the subhalo is then the descendant candidate with the highest score. Finally, the halo finder may not detect a small subhalo that is passing through a larger structure in the subsequent snapshot, because the density contrast is not high enough. Descendants are therefore identified also by skipping one snapshot and considering candidates two snapshots apart.

\subsubsection{SubLink}

{\sc SubLink} constructs merger trees at the subhalo level (see \citealt{rodrig15}), using a merit function equal to the sum of the binding energy ranks of matched particles, raised to a power of $-1$. For handling snapshot skipping, it allows some subhalos to skip a snapshot when finding a descendant. In particular, if the highest ranked descendant two snapshots forward differs from the `descendant of the descendant' found through adjacent snapshots, the former is selected (see Fig. 1 in \citealt{rodrig15}). Once all descendant connections have been made, the main progenitor of each subhalo is defined as the one with the ``most massive history'' behind it (following \citealt{delucia07}).

The {\sc SubLink} merger tree is one large data structure split across several sequential HDF5 files named \linebreak\texttt{tree\_extended.[fileNum].hdf5}, 
where [fileNum] goes from e.g. 0 to 9 for the Illustris-1 run. These files store the data on a per tree basis, and therefore are completely independent from each other. More specifically, any two subhalos that are connected by any of the pointers described in the {\sc SubLink} table are guaranteed to belong to the same tree, and, therefore, their data is found in the same file. Table \ref{table_sublink} lists the fields which are present in each file.

\subsubsection{LHaloTree}

The {\sc LHaloTree} algorithm is virtually identical to that used for the Millennium, Aquarius, and Phoenix simulations, but in HDF5 format. It also constructs trees based on subhalos instead of main halos, and described fully in the supplementary information of \cite{spr05c}. The unique descendant is selected as the subhalo with the highest score, which as before equals the sum of the binding energy ranks of matched particles, raised in this case to a power of $-2/3$. To allow for the possibility that halos may temporarily disappear for one snapshot, the process is repeated for snapshot $n$ to snapshot $n + 2$. If either there is a descendant found in snapshot $n + 2$ but none found in snapshot $n + 1$, or, if the descendant in snapshot $n+1$ has several direct progenitors and the descendant in snapshot $n + 2$ has only one, then a link is made that skips the intervening snapshot. Finally, the main progenitor of each subhalo is selected as the most massive, rather than the one with the most massive history behind it.

The {\sc LHaloTree} merger tree is one large data structure split across several HDF5 files named \linebreak\texttt{trees\_sf1\_135.[chunkNum].hdf5}, 
where [chunkNum] goes from e.g. 0 to 511 for the Illustris-1 run.
Within each file there are a number of groups named ``TreeX'', where X corresponds to the FoF group number in the group catalogs at the final snapshot.
However, note that the number X starts over at zero for each tree file chunk, so the FoF group number is recovered by summing of the number of trees in all previous tree file chunks.
The pair (SubhaloNumber,SnapNum) provides the indexing into the {\sc Subfind} group catalog.
The five other indices for each entry in a TreeX group index into that same group in the tree file. Table \ref{table_lhalotree} describes the fields in the Header and TreeX groups.

\subsection{Supplementary Data Catalogs}

The following additional data products have been computed in post-processing, based on the raw simulation outputs. They are either already available, and now unified under the Illustris data release and made available through the API, or are now made available. In the current effort we focus exclusively on additional properties derived for Illustris-1 galaxies, exclusively at $z=0$ and above a stellar mass limit of $M_\star \gtrsim 10^9 {\rm M}_\odot$.

\subsubsection{Stellar Mocks: Multi-band Images and SEDs}

A catalog of synthetic stellar images and integrated spectra of galaxies in Illustris-1 at $z=0$, produced using the radiative transfer code SUNRISE.
For complete details on this data product, see \cite{torrey15} where it was first described and made available.
For all galaxies with stellar masses $M_\star > 10^{10} {\rm M}_\odot$ ($\sim10^4$ star particles and above), both integrated SEDs and spatially resolved photometric maps in 36 broadband filters are computed. There are approximately 7000 galaxies above this limit. For all galaxies with smaller stellar masses, down to 500 star particles, only integrated SEDs are calculated.
The 36 bands include GALEX, SDSS, IRAC, Johnson, 2MASS, ACS, and preliminary NIRCAM filters.
Note that this is the only data product which is in a format other than HDF5 (namely, FITS).
However, the API provides extractions of individual bands and viewing angles in HDF5 format, as well as SEDs in text format, if requested. Finally, we have developed the Python code {\sc Sunpy}\footnote{\url{http://github.com/ptorrey/sunpy}} to add observational realism and make figures based on the raw stellar mock image FITS files. 

\subsection{Photometric Non-Parametric Stellar Morphologies}

A catalog of photometric non-parametric morphologies of Illustris-1 galaxies at $z=0$.
This is meant to replicate automated diagnostics of galaxy stellar structure commonly used observationally, and is 
calculated by first adding observational realism to the idealized `stellar mock' images from \cite{torrey15}, 
then measuring $(G_{\rm ini}, M_{20}, C, r_P, r_E)$ statistics in four bands, rest-frame u, g, i, and H, each from four directions.
For full details on the calculation of each value, see Table \ref{table_supp_nonpara_morphs} and \cite{snyder15} \citep[following][]{lotz04}.
This data is available for essentially all subhalos with $M_\star > 10^{9.7} {\rm M}_\odot$ at $z=0$ in Illustris-1.
Treating each viewing direction as an independent object, values have been computed for a uniform set of 42531 sources per filter.

\subsection{Stellar Circularities, Angular Momenta, Axis Ratios}

A catalog for the circularities, angular momenta and axis ratios of the stellar component, for Illustris-1 galaxies.
Data is available for all subhalos with stellar mass (inside twice the stellar half mass radius) bigger than $10^9 {\rm M}_\odot $.
For complete definitions on the calculation of each value, see Table \ref{table_supp_stellar_circs} and \cite{genel15}, where they were presented and used.
The first four quantities in Table C.2 are calculated after alignment with the angular momentum vector of the stars within 10 times the stellar half-mass radius, and measure the quantities inside that radius.
The ``Circ*'' fields are based on the distribution of the circularity parameter $\epsilon$ of the individual stars, 
as defined in Equation (1) of \cite{marinacci14a}. Finally, an analogous calculation including the full stellar content of the subhalos is also provided.

%----------------------------------------------------------------
% Data Access
%----------------------------------------------------------------

\section{Data Access} \label{sDataAccess}

There are two complementary ways to access the Illustris data products.

\begin{enumerate}
\item Raw files can be directly downloaded, and example scripts are provided as a starting point for local analysis.
\item A web-based API can be used, either through a web browser or programmatically in an analysis script, to perform common search and extraction tasks.
\end{enumerate}

These two approaches can be combined. For example, a user may be forced to download the full redshift zero group catalog in order to perform a complex search not supported by the API. After locally determining a sample of interesting galaxies, one could then extract their individual merger trees (and/or raw particle data) without needing to download the full simulation merger tree (or a full snapshot). 

Both approaches are documented below, while ``getting started'' tutorials for several languages (currently: Python, IDL, and Matlab) can be found online.

\subsection{Direct File Download and Example Scripts}

All of the primary data products for Illustris are released in HDF5 format.
This is a portable, self-describing, binary specification suitable for large numerical datasets, for which 
file access routines are available in all common computing languages.
We use only the basic features of the format: groups, attributes, and datasets, with one and two dimensional numeric arrays.

In order to maintain reasonable filesizes, most outputs are split across multiple file ``pieces'' (or ``chunks'').
For example, each snapshot of Illustris-1 is split into 512 sequentially numbered files.
Individual links to each file chunk are available through the web-based API, and a snapshot can be downloaded in its entirety with a single {\small wget} command.
Direct download links for other snapshots, simulations, and file types (such as group catalogs or merger trees) can be found at the appropriate URLs, as described below.
Pre-computed {\small sha256} checksums are provided for all files so that their integrity can be verified.

The provided example scripts (in IDL, Python, and Matlab) give basic I/O functionality such as: 
(i) reading a given particle type and/or data field from the snapshot files, 
(ii) reading only the particle subset from the snapshot corresponding to a halo or subhalo, 
(iii) extracting the full subtree or main progenitor branch from either {\sc SubLink} or {\sc LHaloTree} for a given subhalo, 
(iv) walking a tree to count the number of mergers, 
(v) reading the entire group catalog at one snapshot, 
(vi) reading specific fields from the group catalog, or the entries for a single halo or subhalo. 
We expect they will serve as a useful starting point for writing any analysis task, and intend them as a `minimal working examples' which are short and simple enough that they can be quickly understood and extended.

\subsection{Web-based API}

We have implemented a web-based interface (API) which can respond to a variety of user requests and queries. 
It is a well-defined interface between the user and the Illustris data products, which is expressed in terms of the required input(s) and expected output(s) for each type of request. 
The provided functionality is independent, as much as possible, from the underlying data structure, heterogeneity, format, and access methods. 
The API can be used in addition to, or in place of, the download and local analysis of large data files.
At a high level, the API allows a user to \textbf{search}, \textbf{extract}, \textbf{visualize}, and \textbf{analyze}.
In each case, the goal is to reduce the data response size, either by extracting an unmodified subset, or by calculating a derivative quantity.

By specific example, the following types of requests can be handled through the current API, for any simulation at any snapshot:

\begin{itemize}[leftmargin=2em]
\setlength\itemsep{0em}
\item List the available simulations, their snapshots, and all associated metadata.
\item List all objects in the {\sc Subfind} group catalog and their properties.
\item Search with numeric range(s) over any field(s) present in the {\sc Subfind} group catalogs.
\item Return all fields from the group catalog for a specific halo or subhalo.
\item Return a full snapshot cutout of the particle/cell data for a given halo or subhalo.
\item Return a subset of this `group cutout' containing only specified particle/cell type(s), and/or specific field(s) for each type.
\item Return the complete merger history, or just the main progenitor branch, for a given subhalo, for any of the merger trees.
\item Download all raw snapshot, group catalog, merger tree, and supplementary data catalog files which exist.
\item Download subsets of raw snapshot files, containing only specified particle/cell type(s), and/or specific field(s) for each type.
\item Crossmatch subhalos between full physics runs and their dark matter only analogues.
\item Traverse relationships between halos and subhalos, for instance from a satellite subhalo to its parent FoF group to the primary (central) subhalo of that group.
\item Traverse descendant and primary progenitor links across adjacent snapshots, as available in the {\sc SubLink} merger trees.
\item View or render visualizations of the different components (e.g. dark matter, gas, stars) of halos and subhalos, when available.
\item Retrieve or calculate additional properties, beyond what is available in the group catalogs, for halos and subhalos, when available.
\end{itemize}

The Illustris data access API is available at the following permanent URL:

\begin{itemize}[leftmargin=2em]
\item \url{http://www.illustris-project.org/api/}
\end{itemize}

Simple Python examples for working with the API are provided in Appendix D. 
We provide a list of endpoints, their descriptions, and return types. All accept only GET requests.
To provide long-term consistency, we anticipate that the API structure described herein will never change.
As additional data products, simulations, tools, and analysis tasks are developed and released, new endpoints will be added.
In order to take advantage of new features as they are introduced, we recommend a user consult the up to date API reference available on the website. Tables \ref{table_api1} and \ref{table_api2} provide descriptions of each currently available endpoint.

\begin{figure*}[htb!]
\centerline{\includegraphics[angle=0,width=7.2in]{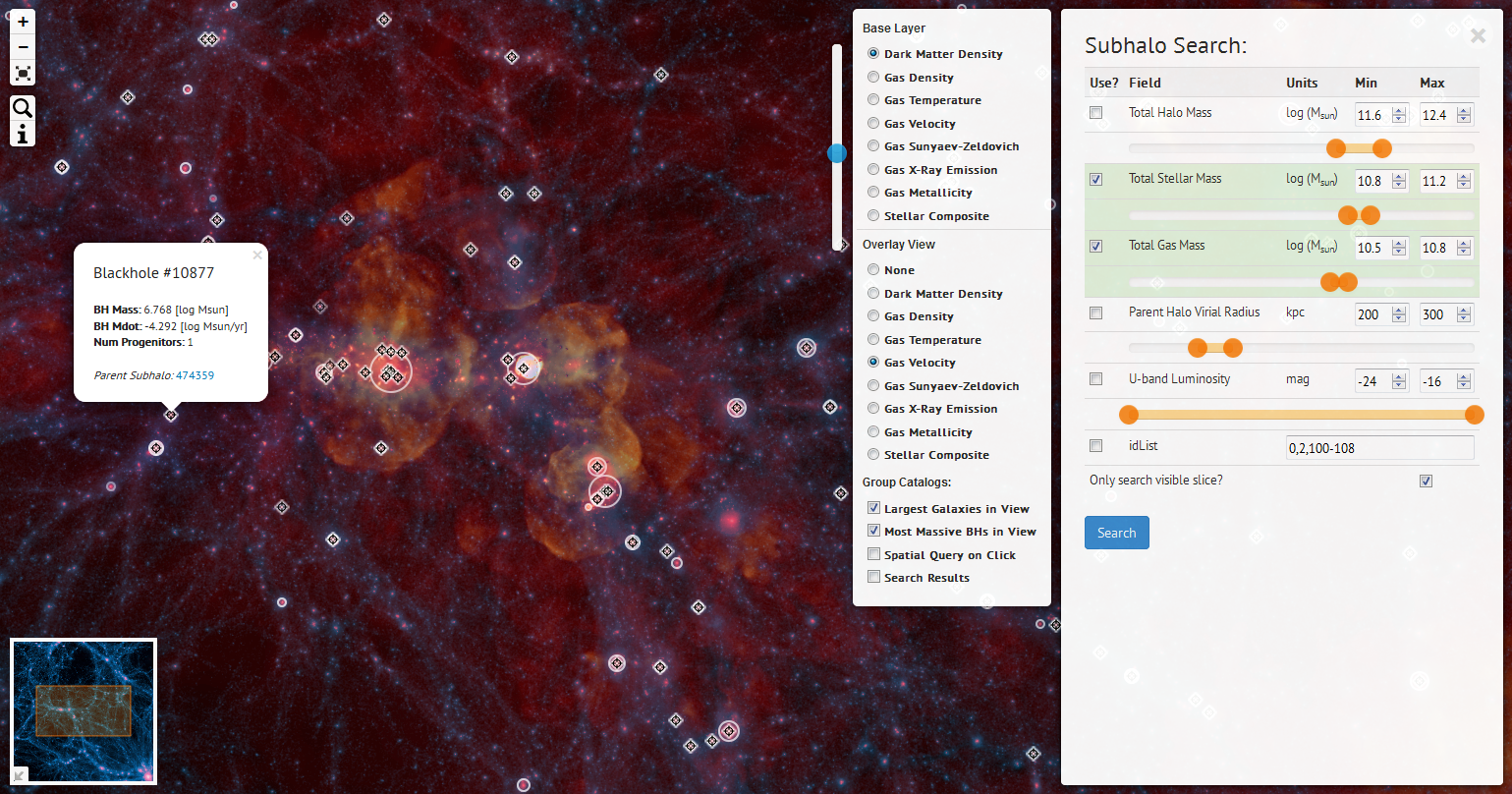}}
\caption{ The current Illustris Explorer interface. The main view shows a gas velocity projection overlaid on the dark matter density field. The most massive galaxies currently visible are shown with circles, while black holes are represented with crosshairs. The overview in the lower left corner provides orientation on larger scales. Clicking at any location will launch a spatial search for the nearest subhalos, while clicking on a BH particle will query its details, including a link to its parent subhalo. The central panel controls image layer selection. The right panel presents a simple search interface over subhalo properties. 
 \label{fig_explorer}} 
\end{figure*}

\subsubsection{API Access Details}

Each API endpoint can return a response in one or more data types. When multiple options exist, a specific return format can be requested through one of the following methods.

\begin{itemize}[leftmargin=2em]
\setlength\itemsep{0em}
\item ``(?format=)'' indicates that the return type is chosen by supplying such a querystring, appended to the URL.
\item ``(.ext)'' indicates that the return type is chosen by supplying the desired file extension in the URL.
\end{itemize}

\noindent \textbf{Search and Cutout Requests.} Several API functions accept additional, optional parameters, which are described here.

\{search\_query\} is an AND combination of restrictions over any of the supported fields, where the 
relations supported are `greater than' (gt), `greater or equal to' (gte), `less than' (lt), `less than or equal to' (lte), `equal to'. 
The first four work by appending e.g. `\_\_gt=val' to the field name (using a double underscore). For example:
 
\begin{itemize}[leftmargin=2em]
\setlength\itemsep{0em}
\item mass\_dm\_\_gt=90.0
\item mass\_\_gt=10.0\&mass\_\_lte=20.0
\item vmax\_\_lt=100.0\&len\_\_gas=0\&vmaxrad\_\_gt=20.0
\end{itemize}

\{cutout\_query\} is a concatenated list of particle fields, separated by particle type. 
The allowed particle types are `dm',`gas',`stars',`bhs'. The field names are exactly as in the snapshots (``all'' is allowed).
Omitting all particle types will return the full cutout: all types, all fields.
For example:

\begin{itemize}[leftmargin=2em]
\setlength\itemsep{0em}
\item gas=Masses,Coordinates,Velocities
\item dm=Coordinates\&stars=all
\end{itemize}

\noindent \textbf{Authentication.} All API requests require authentication, and therefore also user registration. Each request must provide, along with the details of the request itself, the unique ``API Key'' of the user making the request.
A user can send their API key in the querystring, by appending it to the URL as:

\begin{itemize}[leftmargin=2em]
\item ?api\_key=d22d1f16b894a0b894ec31
\end{itemize}

A user can alternatively send their API key in HTTP header. This is particularly useful for {\small wget} commands or within scripts (see the API tutorial). Note that if a user is logged in to the website, then requests \textit{from the browser} are automatically authenticated. Navigating the Browsable API works in this way.

\subsection{Further Online Tools}

\subsubsection{Subhalo Search Form}

We provide a simple search form through which users can query the subhalo database. The search capabilities that exist in the API are exposed in a more human-friendly interface, to enable exploration without the need to write code or write URLs by hand. For example, objects can be selected based on total mass, stellar mass, star formation rate, gas metallicity, or size. The output is a familiar spreadsheet type format, which lists properties from the group catalogs. In addition, each subhalo row provides links to a common set of web-based tools for introspection. These include the canonical link to the object within the API, a form for selecting particle types and initiating an extraction of particles from the snapshot, merger tree visualization, and links to pre-rendered images, when available.

\subsubsection{Explorer}

The Illustris Explorer\footnote{\url{www.illustris-project.org/explorer/}} is an experiment in the visualization, exploration, and dissemination of large data sets -- in particular, those generated by large, astrophysical simulations such as Illustris. It uses the approach of thin-client interaction with derived data products, in this case, pre-computed imagery layered under group catalog information. In Figure \ref{fig_explorer} a full box slice of the simulation is shown in projection, with a depth of 15 Mpc/h, revealing a fifth of the total volume of Illustris at $z=0$. All the imagery is rendered and saved as hierarchical image pyramids (see also \cite{overzier13,khandai14,bertin15}), while rapid search over group properties spatially overlays the results within this volume. All mass components of the simulation are present: the continuous gas and dark matter fields, stellar light from individual stars, and black holes. We have found the interface particularly useful in exploring the spatial relationships between these four components and the discrete halos and subhalos identified with substructure finding algorithms.

\subsubsection{Merger Tree}

\begin{figure*}[ht!]
\centerline{\includegraphics[angle=0,width=7.3in]{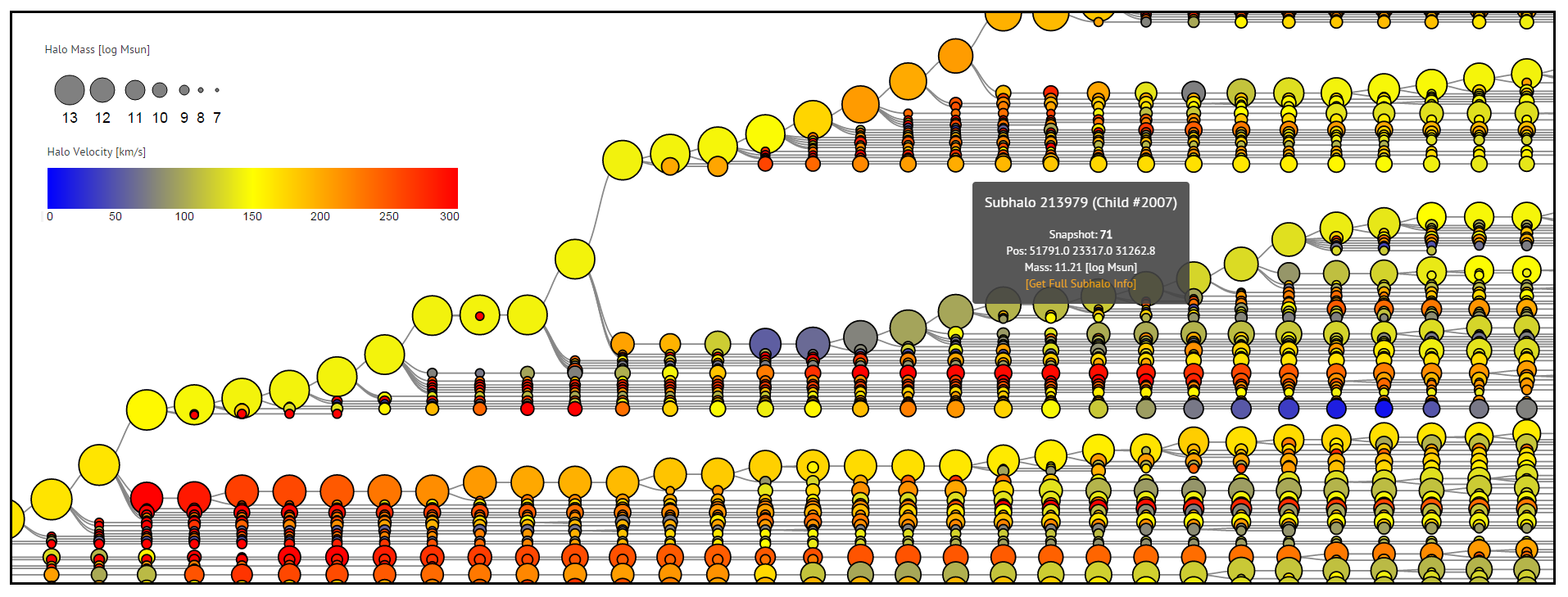}}
\caption{ Example of interactive merger tree exploration. We show a zoomed-in portion of the {\sc SubLink} tree for the 500th most massive central subhalo of Illustris-1 at $z=0$ (ID 395444). Vector based, client-side rendering means that each node can be interacted with individually. One is shown displaying an informational popup, which includes a link back into the API for inspecting that particular progenitor subhalo. Here we show tree node size scaled with total halo mass in $\log {\rm M}_\odot$, and color mapped to subhalo velocity magnitude in $km/s$. 
 \label{fig_mergertree}} 
\end{figure*}

As a demonstration of the potential of rich client applications built on top of the Illustris API, we show in Figure \ref{fig_mergertree} the currently available interface for interactively exploring the merger trees.\footnote{If logged in, this viewer can be launched from inside the Explorer, by selecting a subhalo ID or subhalo circle marker after a search, or through the general subhalo search form.} A zoomed-in portion of the {\sc SubLink} tree for the 500th most massive central subhalo of Illustris-1 at $z=0$ is shown. For any run, snapshot, and subhalo combination, the browser requests a parseable representation of the merger tree from the API (in JSON format), and renders it using the scalable vector graphics (SVG) backend of the d3 javascript visualization library. Because the tree is vector based, and client side, each node can be interacted with individually. Here the informational popup provides a link, back into the API, where the details of the selected progenitor subhalo can be interrogated.

%----------------------------------------------------------------
% Architectural and Implementation Details
%----------------------------------------------------------------

\section{Architectural and Implementation Details} \label{sImplementation}

In the development of the Illustris public data release, many design decisions were made. Here we discuss technical details related to the release effort, focusing on the relationship between (i) expected use cases with preferred methods of data analysis, and (ii) the specific decisions made to enable those goals, balanced against practical considerations and the need for efficiency. We also contrast with other methodologies, as implemented in other large simulation data releases, and attempt to justify the particular balance struck in the case of Illustris. The details in this section are not necessary for scientific uses of the simulation data.

\subsection{Relational Databases}

The vast majority of past simulation data releases have made use of relational database systems (i.e. MySQL, PostgreSQL, or commercial options) as the primary mechanism for user interaction as well as data distribution. Following the impressive success of the SDSS Skyserver \citep{szalay00}, and starting notably for theory with the Millenium simulation database \citep{lemson06}, users were invited to write and submit raw SQL queries to these databases. Most non-trivial tasks require complex queries which can join multiple tables together across foreign key relations, as well as an awareness of the indexing systems and their use. The power of the query language is offset for most non-experts by the unusual approach, which requires abandoning common methods for the local analysis of astronomical data sets: most notably, the writing of small code snippets, which can have loops and if-else type decision branches. Although many science questions relevant for these projects can be answered by writing suitable SQL queries -- as in the ``20 typical queries'' the SDSS system was designed around \citep{gray02} -- it is easy to think up a complex analysis routine which would be unwieldy, if possible at all, with such queries.

In the present effort we have consequently made more limited use of a relational database in the usual way, to hold the full outputs of the group finding algorithms (and not the raw particle data). We exported all group catalogs into the database, with one InnoDB table per run. Each table is partitioned on snapshot number, and has only a single composite B-Tree index on (snapshot,subhalo\_id). The goal was to enable rapid search over arbitrary parameter combinations, primarily at a single snapshot. Therefore we did not adopt a merger tree centric ordering \citep[as in][]{lemson06}. In fact, by releasing multiple merger trees we wished to emphasize the fact that there is no ground truth for the merger history of any object, where by definition such an ordering is useful for only one tree. Our snapshot ordering scheme suffers the same limitation -- it is specifically reflective of the {\sc Subfind} group finder employed on-the-fly. However, based on previous experiences within the collaboration, we have adopted this snapshot ordering scheme as being particularly effective for galaxy-centric analyses. Replication of the particle level data using a different ordering (e.g. along a space-filling curve, as has been typically done) would be prohibitively expensive, and so we offer it only in its single existing format.

For interactions with the group catalogs we hide the existence of the database behind an API facade, instead of allowing the direct submission of SQL queries. This approach implies that each piece of functionality must be exposed through an API endpoint. The trade-offs are clear: common tasks which are supported are much easier to accomplish, while more complex or specialized queries are simply not possible. Our motivations for this decision arose out of several considerations.

First, the complexity of hydrodynamical simulation data, as opposed to the dark matter only case, is substantially higher. The number of properties for each halo or galaxy is larger, and the number of possible analysis and post-processing tasks even more so. Therefore, our expectation from the outset was that users would primarily want to process simulation data using their local computing resources and familiar environments. Given this preference towards local data acquisition and analysis, a large focus of the API is on data volume reduction prior to transfer -- for example, the ability to download particle data for a single galaxy without having to acquire an entire snapshot, or its merger history without having to download the entire merger tree. This is similar in spirit to \cite{rasera10} where particle extraction by halo was also made available, as were sub-volume tilings which together encompass the whole box. Our willingness to promote this approach is driven in part by the increasing availability of high bandwidth network connections, and so the ability to easily download large data volumes. This has undoubtedly also influenced the ``raw data download'' approaches of other recent, large simulation data releases, \cite{skillman14} in particular. One 1.5\,TB full snapshot of Illustris can be downloaded in a little under two days at 10\,MB/s, a realistic goal for U.S. institutional connections. In reality, then, the only prohibitively large data transfer is the entire set of snapshots.

Our second consideration relates to the use of SQL itself. Previous dark matter simulations implementing the ``raw SQL'' approach \citep{lemson06,crocce10,riebe13} demonstrated considerable success in converting users to the language and workflow as a whole, despite it being a relatively unknown tool within the field. The impact of these projects decisively demonstrates the usefulness of this methodology for such projects. Yet, for most users this tool is still foreign, and many uses of the query interface are to simply export data from the database for ingestion into a more familiar data analysis environment. To estimate interest within the community, we conducted an informal survey prior to the design of the Illustris public data release. We report here in brief the most relevant results. Of 125 responses approximately 70\% were graduate students, postdocs, or faculty in the field, evenly split between observers and theorists. Given the wordings of the questions, the majority opinion was that accessing astronomical data sets by writing SQL queries worked ok, but was not their primary choice. Given the options, the favored approaches were search, cutout, and data download interfaces which were programmatically accessible. The least favored options involved writing SQL queries or interacting with temporary storage or intermediate outputs stored on remote servers. For data download, the majority preferred direct download over HTTP, in FITS ($\sim$55\%) or HDF5 ($\sim$35\%) for large binary data and plain text for smaller data sets. We used this input, in combination with our previous experience and the relevant restrictions, to shape the structure of the API and the data release as a whole.

\subsection{API Design and Data Formats}

The Illustris API is based on a representational state transfer architecture \citep[REST, see][]{fielding00}. Requests and responses are transferred over HTTP, and GET is the only supported request verb (meaning that the system is read-only from the user perspective). Individual resources, or ``endpoints'', are identified by their unique URL. The system is stateless, meaning that each request is independent of any previous requests, and must include sufficient information to handle it. The default response type is JSON, a human-readable text format which can be parsed by all modern languages and clients. Because the primary purpose of the API is to serve scientific data sets, HDF5 is chosen as the default response type for binary data. For many resources, the response can be requested in any number of supported formats, which currently include CSV, JSON, HDF5, FITS, PNG, and plain text. All are easily digestable by any modern scripting language, and we consider the exact choices rather unimportant, so long as they are widely supported.

In particular, our choice of HDF5 for the primary data products is driven mainly by practicality -- whatever output format a simulation writes in, and which the simulators therefore interact with for their own science, will be chosen for the broader release. For example, SDF in the case of \cite{skillman14}, or raw binary arrays with metadata in numpy saves for \cite{khandai14}. The only essential requirement is a self-describing binary format, although more sophisticated extraction tasks may be enabled by the features of a specific format. A particularly nice case of this is the use of SDF for direct array slicing through the HTTP protocol (which already supports file subset requests via starting and ending byte positions). Although HDF5 is sufficiently complicated at the bytestream level to make this same approach impossible, our in-memory hyperslab selection method (described below) offers the same functionality with no apparent difference to the user. The only drawback is that responses from the server cannot be blocked (streamed), so the entire requested data set must be temporarily loaded into memory. Given the small size of our community and the expectation of a correspondingly low number of concurrent requests, this has proven to be a non-issue in practice.

The ability of a client to navigate the API and discover available resources is crucial. We generally adopted the principle of Hypermedia as the Engine of Application State (HATEOAS), meaning that users can discover and request resources in the API without needing to know its structure in advance. This is achieved by stating all relationships between objects in terms of the absolute URL at which each object can be found. For example, the final code listing in Appendix D uses the hyperlinked relationship from a given subhalo to its descendant at a different redshift to walk through a merger tree. In addition to the subhalo catalogs, we also export all relevant metadata for simulation runs and snapshots into the database, which enables the overall API structure. In particular, it allows users to freely discover all available resources (e.g. simulations, snapshots, and types of catalogs or particle data available for each) from the common and fixed API root address. This will enable us to seamlessly include new simulations, as well as new data for existing simulations, as later additions to this initial data release.

In terms of the types of interactions with the API, we aim to support only relatively light queries, which the user should anticipate will complete in a few seconds at most. There is no queued or batch query system, where long running queries can be submitted and their progress periodically polled. There is no per-user remote storage (e.g. ``MyDB'', \citealt{li08}). Together, this greatly simplifies the design of the system and maximizes its ease of use, with the implied thought that the typical user workflow will be to download and process specific datasets on their local machine. The ability to offer a remote, persisent, and familiar analysis environment for end users would be a significant though feasible extension of this approach, which we discuss in the following subsection.

As currently designed, users have no need to consider the actual details of where data resides, or how to access it, at the filesystem level. This design goal motivated a system with a split between a front end, which is exposed to the user, and (one or more) back end resources. The separation allows for the two to be in different locations, and for multiple back ends to be supported. In particular, our division is such that the front end handles (i) the Illustris website itself, including (ii) all user details: registration, management, authentication. (iii) All statistics and record keeping. (iv) The full API structure, and responding to API requests at all endpoints. (v) The database, holding both simulation metadata, and the group catalogs. Currently only one back end is in use, and consists of a public-facing machine on the same local network as the data, which is mounted via NFS. It handles:

\begin{itemize}[leftmargin=2em]
\setlength\itemsep{0em}
\item Serving raw data files. In this case, several distributed filesystems are locally mounted. Requests are translated into the appropriate system path, and given back to Apache to serve directly via XSendFile.
\item Extracted subsets of data files are also served. In this case, the pre-calculated offsets are used in order to only read the requested data from disk. This data is either read into a memory structure in the format requested by the client, or subsequently converted to the requested format. In particular, binary extractions from HDF5 containers are read into an in-memory HDF5 ``image''. The raw bytestream of this image is then transferred to the client from memory, such that no temporary copy of the data subset need be saved.
\end{itemize}

The back end is stateless, has no database or persistent local storage of any kind, and no knowledge of the user making each request. This simplifies the addition or transfer of data sources.
In order to provide authentication, which forms the basis of usage monitoring, permission levels, bandwidth throttling and rate limits, the following steps are taken:

\begin{enumerate}[leftmargin=2em]
\item The user makes a request to the API on the front end, including their API-Key.
\item The front end authenticates (verifies their identity) and authorizes (checks sufficient permissions) the user.
\item The front end verifies the validity of the request, including the existence of the requested data.
\item If the request can be satisfied from data available in the front end database (e.g. simulation metadata, subhalo fields), the response is returned directly.
\item If the request requires data from the back end, the appropriate path (URL) is constructed.
\item The front end generates a hash-based message authentication code (HMAC) by concatenating a time-based one-time password (TOTP, see \href{http://tools.ietf.org/html/rfc6238}{RFC 6238}) with a pre-shared secret key and the request URL itself.
\item This token is appended to the back end request URL, which is then sent to the client with a REDIRECT request.
\item The client makes the request to the back end.
\item The back end verifies the request by computing the current TOTP and constructing the same hash using the pre-shared secret key.
\end{enumerate}

The use of the time-varying key means that each request to the back end is attached to a specific request from a specific user. The advantage of this approach is that the front end can redirect clients to data at any back end resource while avoiding the bandwidth burden of making the request itself and forwarding the data on to the client. Although the authentication process is somewhat complex, from the perspective of the user the additional burden is minimal. We find each of its uses important: (i) usage monitoring is needed for our accurate assessement of impact within the community, (ii) different permission levels allow us to include private or pre-release data for specific collaborators within the same framework, while (iii) bandwidth and rate limits can enforce fair use if necessary.

\subsection{Software Stack and Future Directions}

At the software level, the Illustris data release makes use of a large number of projects. It is realized on a common open source software stack: \href{http://www.centos.org/}{CentOS}, \href{http://httpd.apache.org/}{Apache}, and \href{http://www.mysql.com/}{MySQL}. On the front end, \href{http://www.python.org/}{Python} is used to handle all dynamic web content through the \href{http://www.djangoproject.com/}{Django} web framework with several packages including the \href{http://www.django-rest-framework.org/}{Django REST framework}. The website uses the \href{http://getbootstrap.com/}{Bootstrap} framework, the \href{http://jquery.com}{jQuery} javascript library, \href{http://www.mathjax.org}{MathJax} and \href{http://pygments.org/}{pygments} rendering. The Explorer interface uses the \href{http://leafletjs.com/}{Leaflet} tile map engine, as well as the two-dimensional R-Tree indexing capabilities in MySQL to locate subhalos and black holes inside in the visible bounding box. Currently there is no support for spatial indexing in higher dimensions, so using the database for 3D (periodic) distance queries would require a custom solution \citep{lemson11}.

Client-side visualizations, currently for the merger trees, use the \href{http://d3js.org/}{d3} javascript data visualization library, and \href{http://threejs.org/}{three.js} for WebGL. There is significant room for the development of additional features in these areas. In particular, for (i) on-demand visualization tasks, (ii) on-demand analysis tasks, and (iii) client-side, browser based tools for data exploration and visualization. For example, (i) requesting an image of projected gas density for a given halo, (ii) requesting a power-law radial slope measurement of a stellar halo or best-fit NFW parameters, and (iii) an interactive 3D representation of the subhalos within a given halo. We welcome community input and direct contributions in any of these directions. On the back end, the \href{http://www.hdfgroup.org/HDF5/}{HDF5 library} with the \href{http://www.h5py.org/}{h5py}, \href{http://www.numpy.org/}{numpy}, and \href{http://github.com/esheldon/fitsio}{fitsio} Python packages provide the bulk of the data interaction layer.

This back end is currently only focused on storage and data delivery, and we do not yet have any system in place to allow temporary, guest access to compute resources which are local to the data itself. However, we envision that this could change in the future. The data delivery portal has access to the compute resources of the cluster, and instead of defining specific, pre-written analysis functions, we would like to provide a familiar environment for the execution of arbitrary user programs. There has been significant recent development related to remote, multi-user, rich interfaces to computational kernels. In particular, the \href{http://jupyter.org}{Jupyter} notebook environment (previously called IPython, \citealt{perez07}) can be spawned, on demand, inside sand-boxed \href{http://www.docker.com}{Docker} instances, through a web-based portal with authentication provided by the existing user registration system. This means that users could develop analysis routines in any language (Jupyter support includes Python, IDL, Matlab, \href{http://julialang.org/}{Julia}, and many others) and execute them, in the same interface, on the remote cluster. We view this possibility as a promising future direction, particularly for researchers who require such remote resources, and otherwise would be unable to use the data for their science.

Finally, the read-only, highly structured nature of simulation output motivates different and more efficient approaches for data search and processing. As an alternative to search within a relational database, one could consider bitmap indexing over HDF5 as in \href{http://www-vis.lbl.gov/Events/SC05/HDF5FastQuery/}{FastQuery} \citep{chou11,byna12} together with a SQL-like query layer \citep{wang13}. When these technologies are slightly more mature, the need to place a copy of raw simulation data into a database will be removed. Instead, the DB can be used only to handle meta-data, and fast indexed search and queries can be made directly against structured binary data on disk. We anticipate that such an approach might be relevant for future data release efforts, although the sophistication of existing software building blocks already enables an effective way to broadly release both large data sets and rich tools for subsequent data interrogation and analysis.

%----------------------------------------------------------------
% Scientific Remarks and Cautions
%----------------------------------------------------------------

\section{Scientific Remarks and Cautions} \label{sRemarks}

The Illustris Simulations (particularly Illustris-1) have been shown to resolve many details of the small-scale properties of galaxies, as well as the evolution of stars and gas within the cosmic web. Illustris-1 reproduces many observational facts on the demographics and properties of the galaxy populations at various epochs, and on the distribution of gas on large scales. 
As described in Section \ref{sSims}, this has been achieved with a comprehensive galaxy formation model which is intended to account for all the primary processes that are believed to be important for the formation and evolution of galaxies. 

However, the enormous dynamical range and the variety and complexity of physics phenomena involved in these numerical endeavours necessarily involve some modeling uncertainties. 
We have identified below the known problems and points of caution in the Illustris simulated output that any user of the public data must be aware of before embarking on the analysis of the released products. These points should be carefully taken into account before advancing scientific conclusions or making comparisons to observational results.

\subsection{Caveats with the Illustris Galaxy Formation Model}

Limitations in the Illustris implementations of the stellar and AGN feedback, and possibly of the adopted star-formation recipe, determine a series of issues in the simulated galaxy populations and gas content of halos in comparison to observational constraints. These all point to an inefficient quenching of the star formation in galaxies at different masses and regimes, and in some cases also to qualitatively not-realistic behaviors of the feedback models. In particular, we note the following issues applicable to the highest-resolution realization (Illustris-1).

\begin{itemize}[leftmargin=2em]
\setlength\itemsep{0em}
\item The cosmic star formation rate density is too high at $z \lesssim 1$, possibly because of an inefficient quenching of galaxies residing in halos of $10^{11-12} {\rm M}_\odot$ \citep[see Figs. 8 and 2 in][respectively]{vog14b,genel14}.
  
\item The stellar mass function at $z \lesssim 1$ is too high both at the high and the low ends of the sampled stellar mass range, $M_\star \lesssim 10^{10} {\rm M}_\odot$ and $M_\star \gtrsim 10^{11.5} {\rm M}_\odot$, see Fig.11, \cite{vog14b} and Fig.3, \cite{genel14}.

\item The physical extent of galaxies can be a factor of a few larger than observed for $M_\star \lesssim 10^{10.7} {\rm M}_\odot$ \citep[see Fig. 9 in][]{snyder15}.

\item The galaxy color distribution deviates from observations in that it does not exhibit a clear bimodality between red and blue galaxies, and the green-valley and the blue cloud appear over populated with respect to to the red sequence (especially for  $M_\star \gtrsim 10^{10} {\rm M}_\odot$ \citep[see Fig.14 in][]{vog14b}.

\item About 10 percent of disk galaxies in the mass range $M_\star \sim 10^{10.5-11} {\rm M}_\odot$ at $z=0$ exhibit strong stellar and gaseous ring-like features, and appear as an additional sub-population in the $G_{\rm ini}-M_{20}$ plane \citep[see Fig. 5 in][]{snyder15}; such features appear to be even more frequent at higher redshifts. Via fragmentation, stellar rings may give rise to spurious stellar clumps that the {\sc Subfind} algorithm identifies as subhalos but whose origin and existence is not necessarily physically well motivated (see also below). Furthermore, these stellar rings are often associated with cores in the stellar and dark matter components, visible in the inner radial density profiles. These cores can extend up $\sim$\,10 kpc in radius and are likely not realistic in detail.

\item The total gas within $R_{\rm 500c}$ is underestimated at late times by a factor 3-10 in halos with $M_{\rm 500c} \sim 10^{13-14} {\rm M}_\odot$, because of the too violent operation mode of the Illustris radio-mode feedback \citep[see Fig. 10 in][]{genel14}.

\item For similar reasons, the bolometric X-ray luminosity in the hot coronae of elliptical galaxies is by many factors lower than in spiral galaxies, contradicting observational constraints \citep[see Section 5.2 of][]{bogdan15}; and the predictions for the Sunyaev-Zel'dovich signals from Illustris clusters are not reliable (Popa et al. 2015, in prep).
\end{itemize}

For some items of this list we have intentionally omitted more specific quantifications of the tensions with observations for two reasons: on the one side, not all observational results are in agreement among each other, making quantitative statements necessarily partial; on the other side, excruciating care is necessary to properly map simulated variables into observationally-derived quantities. 
For example, we notice that the adopted low star-formation density threshold value and the low thermal energy content of galactic winds may be the cause for spurious star-formation in the circumgalactic medium around Milky Way-like galaxies, at large distances from the natural, dense sites of star formation activity (i.e. disks, see \citealt{marinacci14a}). However, no observational data are available to properly quantify such phenomenon. Similarly, the impact of the AGN feedback on the dark-matter distribution within Illustris halos might be overestimated, but direct observational constraints are lacking.
Furthermore, while a first analysis of the stellar ages of Illustris galaxies seemed to reveal an overestimation of the predicted stellar ages for $M_\star \lesssim 10^{10.5} {\rm M}_\odot$ galaxies (see Fig. 25, \citealt{vog14b}), we have now recognized that such a comparison to observations is rather inconclusive, as the shape of the age-mass relation of galaxies strongly depends, in the first place, on whether stellar ages are measured by mass- or light- weighting.

To better inform which features of the simulations should be trusted when making science conclusions, we note also following points more directly related to numerical choices:

\begin{itemize}[leftmargin=2em]
\setlength\itemsep{0em}
\item In both the snapshots and halo catalogs, metallicity values should be used and interpreted with care. These depend on the underlying choices for stellar evolution and metal enrichment, with tabulated yields being uncertain and continuously updated. Furthermore, no metallicity floor has been imposed to the output data, so that metallicities of a small fraction of gas and star elements adopt minuscule, unrealistic values. In this case, a convenient and appropriate metallicity floor can be adopted, as necessary.

\item In the {\sc Subfind} catalogs, relatively-low mass, stellar- or gas-dominated objects at small galactocentric distances from their host halos may be artifacts and should be considered with care. These may be the results of the fragmentation of aforementioned stellar rings in disk galaxies, and may appear as outliers in halos/galaxies scaling relations involving sizes, masses, metallicities and mass-to-light ratios.

\item Low-mass BHs in relatively low-mass subhalos should also be considered with care, particularly those hosted in satellite subhalos of more massive galaxies or at low redshifts. Because spurious motions of BH particles are prevented by repositioning the BH on halo potential minimum, in some cases, low-mass BHs in satellite galaxies are repositioned on the central halo on artificially short timescales. These ``empty'' satellites may then be repopulated with new BH seeds, regardless of redshift. The vast majority of these late-forming, satellite-hosted seeds do not grow significantly before merging with the central BH, so the effects are largely confined to BHs with mass $<10^6 M_{\odot}$.
\end{itemize}

%----------------------------------------------------------------
% Community Considerations
%----------------------------------------------------------------

\section{Community Considerations} \label{sCommunity}

\subsection{Citation}

To support proper attribution, recognize the effort of individuals involved, and monitor ongoing usage and impact, we request the following.
Any publication making use of data from the Illustris simulations should cite this release paper (Nelson et al. 2015b) as well as the original paper introducing the project \citep{vog14a}.
Furthermore, extensive use of the data, or studies of galaxy properties and populations, should cite if appropriate \cite{vog14b} as well as \cite{genel14}.
Any investigation of the black hole population should cite if appropriate \cite{sijacki14}.

Finally, use of any of the supplementary data products should include the relevant citation. A full and up to date list is 
maintained on the Illustris website. At the time of publication, this includes use of the {\sc SubLink} merger trees \citep{rodrig15},
the redshift zero synthetic stellar images \citep{torrey15}, the subsequently derived morphological parameters \citep{snyder15}, 
and the stellar angular momentum, circularity measurements, and axis ratios \citep{genel15}.

\subsection{Collaboration and Contributions}

The full snapshots of Illustris-1 are sufficiently large that it will be prohibitive for most users to acquire or store a large number. As a result, projects which require access to the entire snapshot set may benefit from closer interaction with members of the Illustris collaboration. In particular, many team members are open to more direct collaboration, which can include guest access to compute resources which are local to full copies of the data. We welcome ideas for joint projects, so long as they intersect with the interests of collaboration members and do not overlap with existing efforts. We suggest, practically, to contact the author(s) who have already published work using Illustris data in related scientific topics.\footnote{See \url{http://www.illustris-project.org/results/} for a list.}

We also welcome contributions to the data release. These can take the form of either analysis code, or computed data products. For example, with the development of an (expensive) analysis routine, we can run it against one or all simulations or snapshots. The resulting data can be made immediately public through the Illustris API. Alternatively, the resulting data can be made privately available until an initial publication is released, and then released publicly. With the development of an (inexpensive, fast) analysis routine, we can integrate it into the Illustris API, such that it can be requested on demand for any object. In this case, analysis should be restricted to subhalo or halo particles, and take at most a few seconds. For the production of a data set derived from the Illustris simulations, in order to make it publicly available, we can host and distribute it alongside the other supplementary data catalogs.

\subsection{Future Data Releases}

We anticipate release of additional data in the near future, for which further documentation will be provided online.

\subsubsection{Rockstar and Consistent-Trees}

We plan to release {\sc Rockstar} group catalogs and the {\sc Consistent-Trees} merger trees built upon them for the six Illustris boxes in the near future, and will provide further documentation at that time. These group catalogs can include a different subhalo population than identified with the {\sc Subfind} algorithm, particularly during mergers. The algorithm used to construct the C-Trees also has fundamental differences to both {\sc LHaloTree} and {\sc SubLink}, inserting `ghost' nodes or modifying properties of existing nodes such that objects in the tree may not map 1-to-1 to the group catalogs from which they were constructed. The output format and structure also differ substantially from either of the two other trees.

These additional catalogs can provide a powerful comparison and consistency check for any scientific analysis. We also anticipate that some users will simply be more familiar with these outputs, or need them as inputs to other tools.

\subsubsection{Additional Supplementary Data Catalogs}

The $z=0$ ``stellar mocks'' multi-band images are being generated for twelve additional snapshots of Illustris-1 at $0.5 < z < 9$. These will include two sets of mock images in 47 common filters, one observing galaxies redshifted to the appropriate epoch and the other observing galaxies in their rest frame. In addition, we expect to add maps of mass, metallicity, gas and stellar velocity, and gas and stellar velocity dispersion in the same projections as these synthetic images.  Subsequently, we will also release the non-parametric morphology catalogs for the high redshift galaxy populations.

We expect to release a mock strong lensing catalog, which includes properties of galaxies that most resemble the observed lenses in term of mass/velocity dispersion.The following properties will be available: the Einstein radius $R_E$, the projected and 3d radial profile slopes, dark matter fraction within $R_E$, central stellar velocity dispersion, anisotropic parameters, effective radius, Sersic index, light ellipticity and orientation. This data will be available at several redshifts from $z=0$ to $z=1$, assuming fiducial source redshifts \citep{xu15}.

Additional details on the black holes will be provided: high time resolution outputs of black hole properties, and enumeration of all black hole merger events. This data is new and independent from the snapshots \citep{blecha15}.

Stellar assembly and merger history catalogs will be released, including details such as in-situ/ex-situ fractions, stellar mass formed pre/post infall, number of major and minor mergers in different time intervals and time since recent merger events. This data will be available for all subhalos at all snapshots of Illustris-123.

Dark-matter halo catalogs at selected snapshots will be released including dark-matter density profiles fit parameters, fit-independent concentration estimates, halo formation times, and halo shapes.

Mock images and property catalogs of Illustris-1 stellar halos will be released, at a selection of snapshots between z=0 to z=2.

We plan to publish lightcone images, whereby we transform raw simulation data from all snapshots into self-consistent mock-observed survey fields, in HST and JWST filters.

\subsubsection{Additional Simulations}

Several smaller simulations related to Illustris have been discussed in previous papers, including a series of $25 {\rm Mpc}/h$ boxes with variations on the input feedback parameters. These can be released in the future if there is community interest. Ongoing and future projects, including higher resolution ``zooms'' of individual systems, as well as larger volumes, will also be released through this platform in the future.

%----------------------------------------------------------------
% Summary and Conclusions
%----------------------------------------------------------------

\section{Summary and Conclusions} \label{sConclusions}

We have made publicly available all the simulated data associated with the Illustris project at the permanent URL:

\begin{itemize}
\item \url{http://www.illustris-project.org/data/}
\end{itemize}

The Illustris project includes a series of large-scale, cosmological simulations ideal for studying the formation and evolution of galaxies. 
The simulation suite consists of three runs at increasing resolution levels of the same (106.5 Mpc)$^3$ cosmological volume, with and without baryonic physics included. 
The high-resolution simulations (Illustris-1 and Illustris-1-Dark) include several million gravitationally bound structures, 
and the $z=0$ Illustris-1 volume contains $\sim$7000 well-resolved galaxies with stellar mass exceeding $10^{10} {\rm M}_\odot$. 
The galaxies sampled in this volume span a range of environments and formation histories, allowing for a wide range of science topics to be addressed using the simulation data.
For all six realizations, we are releasing the following data products:

\begin{itemize}[leftmargin=2em]
\setlength\itemsep{0em}
\item the raw snapshots at all 136 available redshifts down to $z=0$;
\item the friends-of-friends and {\sc Subfind} halo/galaxy catalogs at the same 136 available redshifts down to $z=0$;
\item the {\sc SubLink} and {\sc LHaloTree} merger trees;
\item the raw snapshots of four sub regions of the full volume, for each full physics run, output with significantly higher time frequency;
\item supplementary data catalogs currently focused on properties of the Illustris-1 $z=0$ galaxy population.
\end{itemize}

We anticipate release of additional data post-processed products in the near future, for which further documentation will be provided online. 
Although the total data volume associated with the Illustris project which is presently released is sizeable, $\sim$265 TB, 
we have made a significant effort to make this data accessible to the broader community. 
Specifically, the simulation data is available either via direct download of the raw files or via web-based API queries for common search, extraction, and analysis tasks. 
Extensive documentation on the format and contents of all released datasets is included both in this paper as well as online, where it will be progressively extended. 
Additionally, we have made basic I/O scripts and starting examples in IDL, Python, and Matlab available to enable users to analyze and work with the raw data.
The resulting data products have widespread applications and provide a powerful tool for the interpretation of extragalactic observations.
By making this data publicly available, we hope to maximize the scientific return from the considerable computational resources invested into running the Illustris simulation suite.

%----------------------------------------------------------------
% Acknowledgements
%----------------------------------------------------------------

\section*{Acknowledgements} 

DN would like to thank Research Computing and the Odyssey cluster at Harvard University for significant computational resources.
AP acknowledges support from the HST grant HST-AR-13897.
SG acknowledges support provided by NASA through Hubble Fellowship grant HST-HF2-51341 001-A awarded by the STScI, which is operated by the Association of Universities for Research in Astronomy, Inc., for NASA, under contract NAS5-26555.
VS acknowledges support by the European Research Council under ERC-StG grant EXAGAL-308037, and by the DFG Priority Program SPPEXA through project EXAMAG.
PT acknowledges support from NASA ATP Grant NNX14AH35G.
GS acknowledges support from HST grants HST-AR-12856.01-A and HST-AR-13887.004-A. Funding for HST programs \#12856 and \#13887 is provided by NASA through grants from STScI.
LB acknowledges support provided by NASA through Einstein Fellowship grant PF2-130093.
LH acknowledges support from NASA grant NNX12AC67G and NSF grant AST-1312095.
The authors would like to thank many people for contributing to analysis and understanding of the Illustris simulations and their results: Andreas Bauer, Simeon Bird, Akos Bogdan, Aaron Bray, Eddie Chua, Benjamin Cook, Chris Hayward, Rahul Kannan, Luke Kelley, Cristina Popa, Kevin Schaal, Martin Sparre, Joshua Suresh, Sarah Wellons. 

The Illustris-1 simulation was run on the CURIE supercomputer at CEA/France as part of PRACE project RA0844, and the SuperMUC computer at the Leibniz Computing Centre, Germany, as part of GCS-project pr85je. The further simulations were run on the Harvard Odyssey and CfA/ITC clusters, the Ranger and Stampede supercomputers at the Texas Advanced Computing Center through XSEDE, and the Kraken supercomputer at Oak Rridge National Laboratory through XSEDE.

%----------------------------------------------------------------
% References
%----------------------------------------------------------------

\section*{References}

\bibliographystyle{model2-names}
\bibliography{refs}

%----------------------------------------------------------------
% Appendices
%----------------------------------------------------------------

\clearpage
\appendix
\onecolumn

\section*{Appendix A: Snapshot Data Details}

\setcounter{table}{0}
\renewcommand*{\theHtable}{A.\arabic{table}}
\gdef\thetable{A.\arabic{table}}

\begin{table*}[htb!]
\footnotesize
  \caption{Details on the file organization for the six runs. In each case, $N_f$ represents the number of files for each data type, while the provided sizes are the average for that data type. The approximate total data volume for each run is also listed.}
  \label{table_chunks}
  \begin{center}
  \renewcommand{\arraystretch}{1.5}
    \begin{tabular}{lrccccc}
    \hline
Run & Total $N_{\rm DM}$ & Snapshot $N_f$ & Groupcat $N_f$ & Snapshot Size & Groupcat Size & Data Volume \\ \hline\hline
Illustris-3           & 94,196,375         & 32    & 2     & 22 GB         & 100 MB & 3 TB \\
Illustris-3-Dark      & 94,196,375         & 8     & 2     & 3.2 GB        & 50 MB  & 0.4 TB \\
Illustris-2           & 753,571,000        & 256   & 4     & 176 GB        & 500 MB & 24 TB \\
Illustris-2-Dark      & 753,571,000        & 32    & 4     & 26 GB         & 320 MB & 3.5 TB \\
Illustris-1           & 6,028,568,000      & 512   & 8     & 1.5 TB        & 3.6 GB & 204 TB \\
Illustris-1-Dark      & 6,028,568,000      & 128   & 8     & 203 GB        & 4 GB   & 28 TB \\ \hline
    \end{tabular}
  \end{center}
\end{table*}

\begin{table*}[htb!]
\footnotesize
  \caption{Details of the Header group in the snapshot files.}
  \label{table_snap_header}
  \begin{center}
\renewcommand{\arraystretch}{1.5}
    \begin{tabular}{lccp{9cm}}
    \hline
Field & Dimensions & Units & Description \\ \hline\hline
BoxSize                  & 1 & ${\rm ckpc}/h$ & Spatial extent of the periodic box (in comoving units). \\
MassTable                & 6 & $10^{10} {\rm M}_\odot/h$ & Masses of particle types which have a constant mass (only DM). \\
NumPart\_ThisFile        & 6 & - & Number of particles (of each type) included in this (sub-)file. \\
NumPart\_Total           & 6 & - & Total number of particles (of each type) in this snapshot, modulo $ 2^{32} $. \\
NumPart\_Total\_HighWord & 6 & - & Total number of particles (of each type) in this snapshot, divided by $ 2^{32} $ and rounded downwards. \\
Omega0                   & 1 & - & The cosmological density parameter for matter. \\
OmegaLambda              & 1 & - & The cosmological density parameter for the cosmological constant. \\
Redshift                 & 1 & - & The redshift corresponding to the current snapshot. \\
Time                     & 1 & - &  The scale factor $a =1/(1+z)$ corresponding to the current snapshot. \\
NumFilesPerSnapshot      & 1 & - & Number of file chunks per snapshot. \\ \hline
    \end{tabular}
  \end{center}
\end{table*}

\begin{table*}[htb!]
\footnotesize
  \caption{Additional details of the subbox snapshots. For each subbox number, its physical environment, matter overdensity, center position, box size along each coordinate axis, and volume fraction with respect to the full box.}
  \label{table_subbox2}
  \begin{center}
\renewcommand{\arraystretch}{1.5}
    \begin{tabular}{cccccc}
    \hline
Subbox \# & Environment & $\Omega_m^{\rm sub}$ & $(x_c,y_c,z_c)$ & $L_{\rm subbox}$ & Volume Frac \\ \hline\hline
0 &  Crowded, one $\sim 5\times 10^{13} {\rm M}_\odot$ halo &  1.47 &  (9000, 17000, 63000)  &  7.5 cMpc/$h$ &  0.1\% \\
1 &  Less crowded, several $>10^{12} {\rm M}_\odot$ halos  &  0.16 &  (43100, 53600, 60800) &  8.0 cMpc/$h$ &  0.12\% \\
2 &  Less crowded, several $>10^{12} {\rm M}_\odot$ halos  &  0.29 &  (37000, 43500, 67500) &  5.0 cMpc/$h$ &  0.03\% \\
3 &  Least crowded, several $\sim10^{12} {\rm M}_\odot$ halos &  0.25 &  (64500, 51500, 39500) &  5.0 cMpc/$h$ &  0.03\% \\ \hline
    \end{tabular}
  \end{center}
\end{table*}

\begin{table*}[htb!]
\footnotesize
  \caption{Listing of all snapshot fields for gas (PartType0).}
  \label{table_gas}
  \begin{center}
\renewcommand{\arraystretch}{1.5}
    \begin{tabular}{p{2cm}cp{1.5cm}p{11cm}}
    \hline
Field & Dimensions & Units & Description \\ \hline\hline
Coordinates
& N,3
& $ {\rm ckpc}/h $
& Spatial position within the periodic box of size 75000 {\rm ckpc}/h. Comoving coordinate.
\\
 Density
& N
& $ \frac{ 10^{10} {\rm M}_\odot/h }{ ({\rm ckpc}/h)^3 } $
& Comoving mass density of cell (calculated as mass/volume).
\\
 ElectronAbundance
& N
& -
& Fractional electron number density with respect to the total hydrogen number density, so $n_e = \rm{ElectronAbundance} * n_H$ where $n_H = X_H * \rho / m_p$. Use with caution for star-forming gas (see comment below for NeutralHydrogenAbundance).
\\
 GFM\_\newline AGNRadiation
& N
& $ {\rm erg} / {\rm s} / {\rm cm}^{2} $
& Bolometric intensity (physical units) at the position of this cell arising from the radiation fields of nearby AGN.
\\
 GFM\_\newline CoolingRate
& N
& $ {\rm erg} {\rm cm}^{3} / s $
& The instantaneous net cooling rate experienced by this gas cell, in cgs units (e.g. $ \Lambda_{\rm net} / n_H^2 $).
\\
 GFM\_\newline Metallicity
& N
& -
& The ratio $ M_Z / M_{total} $ where $ M_Z $ is the total mass all metal elements (above He). This is not in solar units! To convert to solar metallicity, divide by 0.0127 (the primordial solar metallicity).
\\
 GFM\_\newline WindDMVelDisp
& N
& $ {\rm km/s} $
& Equal to SubfindVelDisp.
\\
 InternalEnergy
& N
& $ ({\rm km/s})^2 $
& Internal (thermal) energy per unit mass for this gas cell.
\\
 Masses
& N
& $ 10^{10} {\rm M}_\odot/h $
& Gas mass in this cell. Refinement/derefinement attempts to keep this value within a factor of two of the targetGasMass for every cell.
\\
 Neutral\newline Hydrogen\newline Abundance
& N
& -
& Fraction of the hydrogen cell mass (or density) in neutral hydrogen, so $ n_{H_0} = \rm{NeutralHydrogenAbundance} * n_H $. (So note that $n_{H^+} = n_H - n_{H_0}$). Use with caution for star-forming gas, as the calculation is based on the 'effective' temperature of the equation of state, which is not a physical temperature.
\\
 NumTracers
& N
& -
& The number of child tracers residing within this gas cell.
\\
 ParticleIDs
& N
& -
& The unique ID (uint64) of this gas cell. Constant for the duration of the simulation. May cease to exist (as gas) in a future snapshot due to conversion into a star/wind particle, accretion into a BH, or a derefinement event.
\\
 Potential
& N
& $ ({\rm km/s})^2 $
& Gravitational potential energy.
\\
 SmoothingLength
& N
& $ {\rm ckpc}/h $
& Twice the maximum radius of all Delaunay tetrahedra that have this cell at a vertex in comoving units ($s_i$ from Springel et al. 2010).
\\
 StarFormationRate
& N
& $ {\rm M}_\odot / {\rm yr} $
& Instantaneous star formation rate of this gas cell. 
\\
 SubfindDensity
& N
& $ \frac{ 10^{10} {\rm M}_\odot/h }{ ({\rm ckpc}/h)^3 } $
& The local total comoving mass density, estimated using the standard cubic-spline SPH kernel over all particles/cells within a radius of SubfindHsml.
\\
 SubfindHsml
& N
& $ {\rm ckpc}/h $
& The comoving radius of the sphere centered on this cell enclosing the 64$\pm$1 nearest dark matter particles.
\\
 SubfindVelDisp
& N
& $ {\rm km/s} $
& The 3D velocity dispersion of all dark matter particles within a radius of SubfindHsml of this cell.
\\
 Velocities
& N,3
& $ {\rm km} \sqrt{a} / {\rm s} $
& Spatial velocity. The peculiar velocity is obtained by multiplying this value by $ \sqrt{a} $.
\\
 Volume
& N
& $ 1 / ({\rm ckpc}/h)^3 $
& Comoving volume of the Voronoi gas cell. \\ \hline
    \end{tabular}
  \end{center}
\end{table*}

\begin{table*}[htb!]
\footnotesize
  \caption{Listing of all snapshot fields for dark matter (PartType1).}
  \label{table_dm}
  \begin{center}
\renewcommand{\arraystretch}{1.5}
    \begin{tabular}{p{2cm}cp{1.5cm}p{11cm}}
    \hline %\rule{0pt}{2.5ex} 
Field & Dimensions & Units & Description \\ \hline\hline
Coordinates
& N,3
& $ {\rm ckpc}/h $
& Spatial position within the periodic box of size 75000 {\rm ckpc}/h. Comoving coordinate.
\\
 ParticleIDs
& N
& -
& The unique ID (uint64) of this DM particle. Constant for the duration of the simulation.
\\
 Potential
& N
& $ ({\rm km/s})^2 $
& Gravitational potential energy.
\\
 SubfindDensity
& N
& $ \frac{ 10^{10} {\rm M}_\odot/h }{ ({\rm ckpc}/h)^3 } $
& The local total comoving mass density, estimated using the standard cubic-spline SPH kernel over all particles/cells within a radius of SubfindHsml.
\\
 SubfindHsml
& N
& $ {\rm ckpc}/h $
& The comoving radius of the sphere centered on this particle enclosing the 64$\pm$1 nearest dark matter particles.
\\
 SubfindVelDisp
& N
& $ {\rm km/s} $
& The 3D velocity dispersion of all dark matter particles within a radius of SubfindHsml.
\\
 Velocities
& N,3
& $ {\rm km} \sqrt{a} / {\rm s} $
& Spatial velocity. The peculiar velocity is obtained by multiplying this value by $ \sqrt{a} $. \\ \hline
    \end{tabular}
  \end{center}
\end{table*}

\begin{table*}[htb!]
\footnotesize
  \caption{Listing of all snapshot fields for tracer particles (PartType3).}
  \label{table_tracers}
  \begin{center}
\renewcommand{\arraystretch}{1.5}
    \begin{tabular}{p{2cm}cp{1.5cm}p{11cm}}
    \hline
Field & Dimensions & Units & Description \\ \hline\hline
FluidQuantities
& N,13
& Various
& Thirteen auxiliary quantities stored for each tracer with differing significance. See Tracer Quantities below.
\\
 ParentID
& N
& -
& The unique ID (uint64) of the parent of this tracer. Could be a gas cell, star, wind phase cell, or BH.
\\
 TracerID
& N
& -
& The unique ID (uint64) of this tracer. Constant for the duration of the simulation. \\ \hline
    \end{tabular}
  \end{center}
\end{table*}

\begin{table*}[htb!]
\footnotesize
  \caption{Listing of all snapshot fields for stars (PartType4).}
  \label{table_stars}
  \begin{center}
\renewcommand{\arraystretch}{1.5}
    \begin{tabular}{p{2cm}cp{1.5cm}p{11cm}}
    \hline
Field & Dimensions & Units & Description \\ \hline\hline
Coordinates
& N,3
& $ {\rm ckpc}/h $
& Spatial position within the periodic box of size 75000 {\rm ckpc}/h. Comoving coordinate.
\\
 GFM\_InitialMass
& N
& $ 10^{10} {\rm M}_\odot/h $
& Mass of this star particle when it was formed (will subsequently decrease due to stellar evolution).
\\
 GFM\_Metallicity
& N
& -
& See entry under PartType0. Inherited from the gas cell spawning/converted into this star, at the time of birth.
\\
 GFM\_Stellar \newline FormationTime
& N
& -
& The exact time (given as the scale factor) when this star was formed. \textbf{Note: The only differentiation between a real star ($>=0$) and a wind phase gas cell ($<0$) is the sign of this quantity.}
\\
 GFM\_Stellar \newline Photometrics
& N,8
&  $ mag $
& Stellar magnitudes in eight bands: U, B, V, K, g, r, i, z. In detail, these are: Buser's X filter \citep{buser78}, where X=U,B3,V (Vega magnitudes), then IR K filter + Palomar 200 IR detectors + atmosphere.57 (Vega), then SDSS Camera X Response Function, airmass = 1.3 (June 2001), where X=g,r,i,z (AB magnitudes). They can be found in the filters.log file in the BC03 package\footnote{\url{http://www2.iap.fr/users/charlot/bc2003/}}. The details on the four SDSS filters can be found in \cite{stoughton02}, section 3.2.1.
\\
 Masses
& N
& $ 10^{10} {\rm M}_\odot/h $
& Mass of this star or wind phase cell.
\\
 NumTracers
& N
& -
& Number of child tracers belonging to this star/wind phase cell.
\\
 ParticleIDs
& N
& -
& The unique ID (uint64) of this star/wind cell. Constant for the duration of the simulation.
\\
 Potential
& N
& $ ({\rm km/s})^2 $
& Gravitational potential energy.
\\
 SubfindDensity
& N
& $ \frac{ 10^{10} {\rm M}_\odot/h }{ ({\rm ckpc}/h)^3 }$
& The local total comoving mass density, estimated using the standard cubic-spline SPH kernel over all particles/cells within a radius of SubfindHsml.
\\
 SubfindHsml
& N
& $ {\rm ckpc}/h $
& The comoving radius of the sphere centered on this star particle enclosing the 64$\pm$1 nearest dark matter particles.
\\
 SubfindVelDisp
& N
& $ {\rm km/s} $
& The 3D velocity dispersion of all dark matter particles within a radius of SubfindHsml.
\\
 Velocities
& N,3
& $ {\rm km} \sqrt{a} / {\rm s} $
& Spatial velocity. The peculiar velocity is obtained by multiplying this value by $ \sqrt{a} $. \\ \hline
    \end{tabular}
  \end{center}
\end{table*}

\begin{table*}[htb!]
\footnotesize
  \caption{Listing of all snapshot fields for black holes (PartType5).}
  \label{table_bhs}
  \begin{center}
\renewcommand{\arraystretch}{1.5}
    \begin{tabular}{p{1.5cm}cp{3cm}p{10cm}}
    \hline
Field & Dimensions & Units & Description \\ \hline\hline
BH\_CumEgy \newline Injection\_QM
& N
& \rule{0pt}{3.5ex} $ \frac{ 10^{10} {\rm M}_\odot/h ({\rm ckpc}/h)^2 }{ (0.978 {\rm Gyr}/h)^2 } $
& Cumulative amount of thermal AGN feedback energy injected into surrounding gas in the quasar mode.
\\
 BH\_CumMass \newline Growth\_QM
& N
& $ (10^{10} {\rm M}_\odot/h) $
& Cumulative mass accreted onto the BH in the quasar mode.
\\
 BH\_Density
& N
& $ \frac{ 10^{10} {\rm M}_\odot/h }{ ({\rm ckpc}/h)^3 } $
& Local comoving gas density averaged over the nearest neighbors of the BH.
\\
 BH\_Hsml
& N
& $ {\rm ckpc}/h $
& The comoving radius of the sphere enclosing the 64 nearest-neighbor gas cells around the BH.
\\
 BH\_Mass
& N
& $ 10^{10} {\rm M}_\odot/h $
& Actual mass of the BH, does not include gas reservoir. Monotonically increases with time according to the accretion prescription, starting from the seed mass.
\\
 BH\_Mass\_bubbles
& N
& $ 10^{10} {\rm M}_\odot/h $
& Accreted mass in current duty cycle for AGN radio mode bubble feedback. When this value reaches a critical fraction of BH\_Mass\_ini, the bubble energy is released.
\\
 BH\_Mass\_ini
& N
& $ 10^{10} {\rm M}_\odot/h $
& BH mass at the start of the current duty cycle for AGN radio mode feedback, reset after each duty cycle. See BH\_Mass\_bubbles.
\\
 BH\_Mdot
& N
& $ \frac{ 10^{10} {\rm M}_\odot/h }{ 0.978 {\rm Gyr}/h }$
& The mass accretion rate onto the black hole, instantaneous.
\\
 BH\_Pressure
& N
& \rule{0pt}{2.5ex} $ \frac{ 10^{10} {\rm M}_\odot/h }{ ({\rm ckpc}/h)  (0.978 {\rm Gyr}/h)^2 } $
& Reference gas pressure (in comoving units) near the BH, defined as $ (\gamma-1) \rho_{sfr} u_{eq} $, where $ \rho_{sfr}$ is the star-formation threshold and $u_{eq}$ is BH\_U (defined below).
\\
 BH\_Progs
& N
& -
& Total number of BHs that have merged into this BH.
\\
 BH\_U
& N
& $ ({\rm km/s})^2 $
& Thermal energy per unit mass in quasar-heated bubbles near the BH, assuming equilibrium between radiative cooling and thermal AGN heating near the BH. Used to define the BH\_Pressure. 
\\
 Coordinates
& N,3
& $ {\rm ckpc}/h $
& Spatial position within the periodic box of size 75000 {\rm ckpc}/h. Comoving coordinate.
\\
 HostHaloMass
& N
& $ 10^{10} {\rm M}_\odot/h $
& Mass of FoF group that hosts the BH.
\\
 Masses
& N
& $ 10^{10} {\rm M}_\odot/h $
& Total mass of the black hole particle. Includes the gas reservoir from which accretion is tracked onto the actual BH mass (see BH\_Mass).
\\
 NumTracers
& N
& -
& The number of child tracers residing within this BH.
\\
 ParticleIDs
& N
& -
& The unique ID (uint64) of this black hole. Constant for the duration of the simulation. May cease to exist in a future snapshot due to a BH merger.
\\
 Potential
& N
& $ ({\rm km/s})^2 $
& Gravitational potential at the location of the BH.
\\
 SubfindDensity
& N
& $ \frac{ 10^{10} {\rm M}_\odot/h }{ ({\rm ckpc}/h)^3 } $
& The local total comoving mass density, estimated using the standard cubic-spline SPH kernel over all particles/cells within a radius of SubfindHsml.
\\
 SubfindHsml
& N
& $ {\rm ckpc}/h $
& The comoving radius of the sphere centered on this black hole particle enclosing the 64$\pm$1 nearest dark matter particles.
\\
 SubfindVelDisp
& N
& $ {\rm km/s} $
& The 3D velocity dispersion of all dark matter particles within a radius of SubfindHsml.
\\
 Velocities
& N,3
& $ {\rm km} \sqrt{a} / {\rm s} $
& Spatial velocity. The peculiar velocity is obtained by multiplying this value by $ \sqrt{a} $. \\ \hline
    \end{tabular}
  \end{center}
\end{table*}

\begin{table*}[htb!]
\footnotesize
  \caption{Listing of the thirteen auxiliary values stored by the tracer particles. The Reset column indicates whether or not this field is set to zero immediately after each snapshot is written.}
  \label{table_tracer_quantities}
  \begin{center}
\renewcommand{\arraystretch}{1.5}
    \begin{tabular}{cp{2cm}cp{2cm}p{10cm}}
    \hline
Number & Name & Reset? & Units & Description \\ \hline\hline
0
& TMax
& Y
& Kelvin
& The maximum past temperature of the parent gas cell, back to the previous snapshot. Only updated when parent is a gas cell.
\\
 1
& TMax\_Time
& Y
& -
& Scale factor of the above TMax event. Only updated when parent is a gas cell.
\\
 2
& TMax\_Time\_Rho
& Y
& $ \frac{ 10^{10} {\rm M}_\odot / h }{ ({\rm ckpc}/h)^3 }$
& Density of the parent gas cell when the most recent TMax was recorded. Only updated when parent is a gas cell.
\\
 3
& RhoMax
& Y
& $ \frac{ 10^{10} {\rm M}_\odot / h }{ ({\rm ckpc}/h)^3 }$
& Maximum past density of the parent gas cell, back to the previous snapshot. Only updated when parent is a gas cell.
\\
 4
& RhoMax\_Time
& Y
& -
& Scale factor of the above RhoMax event. Only updated when parent is a gas cell.
\\
 5
& MachMax
& Y
& -
& Maximum past mach number of the parent gas cell, as set in the Riemann solver. Only updated when parent is a gas cell.
\\
 6
& EntMax
& Y
& $ P / (\rho / a^3)^\gamma $
& Maximum past entropy of the parent gas cell, back to the previous snapshot. Only updated when parent is a gas cell. Note slightly strange units, where $P$ and $\rho$ are pressure and density, as in the snapshots.
\\
 7
& EntMax\_Time
& Y
& -
& Scale factor of the above EntMax event. Only updated when parent is a gas cell.
\\
 8
& Last\_Star\_Time
& N
& -
& Scale factor, set only when this tracer exchanges from a star/wind to a gas, or from a gas to a star/wind. These four cases respectively set LST = \{ a, -a, a+1, a+2 \}.
\\
 9
& Wind\_Counter
& N
& int32
& Integer counter initialized to zero, increased by one each time this tracer is moved from a gas cell to a wind particle.
\\
 10
& Exchange\_Counter
& N
& int32
& Integer counter initialized to zero, increased by one each time this tracer is exchanged, regardless of parent type.
\\
 11
& Exchange\_Distance
& N
& $ {\rm ckpc}/h $
& Cumulative sum of the spatial distance over which this tracer has moved due to Monte Carlo exchange between gas cells. In particular, the sum of the parent gas cell radii when either the originating parent or destination parent is of gas type.
\\
 12
& Exchange\_\newline Distance\_Error
& N
& $ {\rm ckpc}/h $
& Cumulative sum of $r_{\rm cell} \times ( \sqrt{N_{\rm exch}} - \sqrt{N_{\rm exch}-1} )$, when either the originating or destination parent is of gas type. \\ \hline
    \end{tabular}
  \end{center}
\end{table*}

\clearpage
\section*{Appendix B: Group and Merger Tree Data Details}

\setcounter{table}{0}
\renewcommand*{\theHtable}{B.\arabic{table}}
\gdef\thetable{B.\arabic{table}}

\begin{table*}[htb!]
\footnotesize
  \caption{Description of all fields in the FoF halo catalogs. All fields are float32 unless otherwise specified.}
  \label{table_gc_fof}
  \begin{center}
\renewcommand{\arraystretch}{1.5}
    \begin{tabular}{lccp{10cm}}
    \hline
Field & Dimensions & Units & Description \\ \hline\hline
GroupBHMass
& N
& $ 10^{10} {\rm M}_\odot / h $
& Sum of the BH\_Mass field of all black holes (type 5) in this group.
\\
 GroupBHMdot
& N
& $ \frac{10^{10} {\rm M}_\odot/h }{ (0.978 {\rm Gyr}/h) }$
& Sum of the BH\_Mdot field of all black holes (type 5) in this group.
\\
 GroupCM
& N,3
& $ {\rm ckpc}/h $
& Center of mass of the group, computed as the sum of the mass weighted relative coordinates of all particles/cells in the group, of all types. Comoving coordinate. (Available only for the Illustris-3 run)
\\
 GroupFirstSub
& N
& -
& Index into the Subhalo table of the first/primary/most massive {\sc Subfind} group within this FoF group (int32).
\\
 GroupGasMetallicity
& N
& -
& Mass-weighted average metallicity (Mz/Mtot, where Z = any element above He) of all gas cells in this FOF group.
\\
 GroupLen
& N
& -
& Integer counter of the total number of particles/cells of all types in this group (int32).
\\
 GroupLenType
& N,6
& -
& Integer counter of the total number of particles/cells, split into the six different types, in this group. Note: Wind phase cells are counted as stars (type 4) for GroupLenType (int32).
\\
 GroupMass
& N
& $ 10^{10} {\rm M}_\odot / h $
& Sum of the individual masses of every particle/cell, of all types, in this group.
\\
 GroupMassType
& N,6
& $ 10^{10} {\rm M}_\odot / h $
& Sum of the individual masses of every particle/cell, split into the six different types, in this group. Note: Wind phase cells are counted as gas (type 0) for GroupMassType.
\\
 GroupNsubs
& N
& -
& Count of the total number of {\sc Subfind} groups within this FoF group (int32).
\\
 GroupPos
& N,3
& $ {\rm ckpc}/h $
& Spatial position within the periodic box of size 75000 {\rm ckpc}/h of the maximum bound particle. Comoving coordinate.
\\
 GroupSFR
& N
& $ {\rm M}_\odot / {\rm yr} $
& Sum of the individual star formation rates of all gas cells in this group.
\\
 GroupStarMetallicity
& N
& -
& Mass-weighted average metallicity (Mz/Mtot, where Z = any element above He) of all star particles in this FOF group.
\\
 GroupVel
& N,3
& $ {\rm km/s}/a $
& Velocity of the group, computed as the sum of the mass weighted velocities of all particles/cells in this group, of all types. The peculiar velocity is obtained by multiplying this value by $1/a.$
\\
 GroupWindMass
& N
& $ 10^{10} {\rm M}_\odot / h $
& Sum of the individual masses of all wind phase gas cells (type 4, BirthTime $<= 0$) in this group.
\\
 Group\_M\_Crit200
& N
& $ 10^{10} {\rm M}_\odot / h $
& Total mass of this group enclosed in a sphere whose mean density is 200 times the critical density of the Universe, at the time the halo is considered.  
\\
 Group\_M\_Crit500
& N
& $ 10^{10} {\rm M}_\odot / h $
& Likewise, but for 500 times the critical density of the Universe.
\\
 Group\_M\_Mean200
& N
& $ 10^{10} {\rm M}_\odot / h $
& Likewise, but for 200 times the mean density of the Universe.
\\
 Group\_M\_TopHat200
& N
& $ 10^{10} {\rm M}_\odot / h $
& Likewise, but for $\Delta_c$ times the critical density of the Universe, where $\Delta_c$ derives from the solution of the collapse of a spherical top-hat perturbation (fitting formula from \cite{bryan98}). The subscript 200 can be ignored.
\\
 Group\_R\_Crit200
& N
& $ {\rm ckpc}/h $
& Comoving radius of a sphere centered at the GroupPos of this Group whose mean density is 200 times the critical density of the Universe, at the time the halo is considered.
\\
 Group\_R\_Crit500
& N
& $ {\rm ckpc}/h $
& Likewise, but for 500 times the critical density of the Universe.
\\
 Group\_R\_Mean200
& N
& $ {\rm ckpc}/h $
& Likewise, but for 200 times the mean density of the Universe.
\\
 Group\_R\_TopHat200
& N
& $ {\rm ckpc}/h $
& Likewise, but for $\Delta_c$ times the critical density of the Universe. \\ \hline
    \end{tabular}
  \end{center}
\end{table*}

\begin{table*}[htb!]
\footnotesize
  \caption{Description of all fields in the {\sc Subfind} subhalo catalogs (Part I). All fields are float32 unless otherwise specified.}
  \label{table_gc_subfind1}
  \begin{center}
\renewcommand{\arraystretch}{1.5}
    \begin{tabular}{p{4cm}ccp{9cm}}
    \hline
Field & Dimensions & Units & Description \\ \hline\hline
SubhaloBHMass
& N
& $ 10^{10} {\rm M}_\odot/h $
& Sum of the masses of all black holes in this subhalo.
\\
 SubhaloBHMdot
& N
& $ \frac{ 10^{10} {\rm M}_\odot/h }{ 0.978 {\rm Gyr}/h }$
& Sum of the instantaneous accretion rates $\dot{M}$ of all black holes in this subhalo.
\\
 SubhaloCM
& N,3
& $ {\rm ckpc}/h $
& Comoving center of mass of the Subhalo, computed as the sum of the mass weighted relative coordinates of all particles/cells in the Subhalo, of all types.
\\
 SubhaloGasMetallicity
& N
& -
& Mass-weighted average metallicity (Mz/Mtot, where Z = any element above He) of the gas cells bound to this Subhalo, but restricted to cells within twice the stellar half mass radius.
\\
 SubhaloGasMetallicityHalfRad
& N
& -
& Same as SubhaloGasMetallicity, but restricted to cells within the stellar half mass radius.
\\
 SubhaloGasMetallicityMaxRad
& N
& -
& Same as SubhaloGasMetallicity, but restricted to cells within the radius of $V_{max}$.
\\
 SubhaloGasMetallicitySfr
& N
& -
& Mass-weighted average metallicity (Mz/Mtot, where Z = any element above He) of the gas cells bound to this Subhalo, but restricted to cells which are star forming.
\\
 SubhaloGasMetallicitySfrWeighted
& N
& -
& Same as SubhaloGasMetallicitySfr, but weighted by the cell star-formation rate rather than the cell mass.
\\
 SubhaloGrNr
& N
& -
& Index into the Group table of the FOF host/parent of this Subhalo (int32).
\\
 SubhaloHalfmassRad
& N
& $ {\rm ckpc}/h $
& Comoving radius containing half of the total mass (SubhaloMass) of this Subhalo.
\\
 SubhaloHalfmassRadType
& N,6
& $ {\rm ckpc}/h $
& Comoving radius containing half of the mass of this Subhalo split by Type (SubhaloMassType).
\\
 SubhaloIDMostbound
& N
& -
& The ID of the particle with the smallest binding energy (could be any type, int64).
\\
 SubhaloLen
& N
& -
& Total number of member particle/cells in this Subhalo, of all types (int32).
\\
 SubhaloLenType
& N,6
& -
& Total number of member particle/cells in this Subhalo, separated by type (int32).
\\
 SubhaloMass
& N
& $ 10^{10} {\rm M}_\odot/h $
& Total mass of all member particle/cells which are bound to this Subhalo, of all types.
\\
 SubhaloMassInHalfRad
& N
& $ 10^{10} {\rm M}_\odot/h $
& Sum of masses of all particles/cells within the stellar half mass radius.
\\
 SubhaloMassInHalfRadType
& N,6
& $ 10^{10} {\rm M}_\odot/h $
& Sum of masses of all particles/cells (split by type) within the stellar half mass radius.
\\
 SubhaloMassInMaxRad
& N
& $ 10^{10} {\rm M}_\odot/h $
& Sum of masses of all particles/cells within the radius of $V_{max}$.
\\
 SubhaloMassInMaxRadType
& N,6
& $ 10^{10} {\rm M}_\odot/h $
& Sum of masses of all particles/cells (split by type) within the radius of $V_{max}$.
\\
 SubhaloMassInRad
& N
& $ 10^{10} {\rm M}_\odot/h $
& Sum of masses of all particles/cells within twice the stellar half mass radius.
\\
 SubhaloMassInRadType
& N,6
& $ 10^{10} {\rm M}_\odot/h $
& Sum of masses of all particles/cells (split by type) within twice the stellar half mass radius. \\ \hline
    \end{tabular}
  \end{center}
\end{table*}

\begin{table*}[htb!]
\footnotesize
  \caption{Description of all fields in the {\sc Subfind} subhalo catalogs (Part II). All fields are float32 unless otherwise specified. Note that for all mass calculations by type, wind phase cells are counted as gas.}
  \label{table_gc_subfind2}
  \begin{center}
\renewcommand{\arraystretch}{1.5}
    \begin{tabular}{p{4cm}ccp{9cm}}
    \hline
Field & Dimensions & Units & Description \\ \hline\hline
SubhaloMassType
& N,6
& $ 10^{10} {\rm M}_\odot/h $
& Total mass of all member particle/cells which are bound to this Subhalo, separated by type. 
\\
 SubhaloParent
& N
& -
& Index into the Subhalo table of the unique {\sc Subfind} parent of this Subhalo (int32).
\\
 SubhaloPos
& N,3
& $ {\rm ckpc}/h $
& Spatial position within the periodic box of size 75000 {\rm ckpc}/$h$ of the maximum bound particle. Comoving coordinate.
\\
 SubhaloSFR
& N
& $ {\rm M}_\odot / {\rm yr} $
& Sum of the individual star formation rates of all gas cells in this subhalo.
\\
 SubhaloSFRinHalfRad
& N
& $ {\rm M}_\odot / {\rm yr} $
& Same as SubhaloSFR, but restricted to cells within the stellar half mass radius.
\\
 SubhaloSFRinMaxRad
& N
& $ {\rm M}_\odot / {\rm yr} $
& Same as SubhaloSFR, but restricted to cells within the radius of $V_{max}$.
\\
 SubhaloSFRinRad
& N
& $ {\rm M}_\odot / {\rm yr} $
& Same as SubhaloSFR, but restricted to cells within twice the stellar half mass radius.
\\
 SubhaloSpin
& N,3
& $ (kpc/h) ({\rm km/s}) $
& Total spin per axis, computed for each as the mass weighted sum of the relative coordinate times relative velocity of all member particles/cells.
\\
 SubhaloStarMetallicity
& N
& -
& Mass-weighted average metallicity (Mz/Mtot, where Z = any element above He) of the star particles bound to this Subhalo, but restricted to stars within twice the stellar half mass radius.
\\
 SubhaloStarMetallicityHalfRad
& N
& -
& Same as SubhaloStarMetallicity, but restricted to stars within the stellar half mass radius.
\\
 SubhaloStarMetallicityMaxRad
& N
& -
& Same as SubhaloStarMetallicity, but restricted to stars within the radius of $V_{max}$.
\\
 SubhaloStellarPhotometrics
& N,8
& $ mag $
& Eight bands: U, B, V, K, g, r, i, z. Magnitudes based on the summed-up luminosities of all the stellar particles of the group. For details on the bands, see snapshot details.
\\
 SubhaloStellarPhotometrics \newline MassInRad
& N
& $ 10^{10} {\rm M}_\odot/h $
& Sum of the mass of the member stellar particles, but restricted to stars within the radius SubhaloStellarPhotometricsRad.
\\
 SubhaloStellarPhotometricsRad
& N
& $ {\rm ckpc}/h $
& Radius at which the surface brightness profile (computed from all member stellar particles) drops below the limit of 20.7 mag arcsec$^{-2}$ in the K band (in comoving units).
\\
 SubhaloVel
& N,3
& $ {\rm km/s} $
& Peculiar velocity of the group, computed as the sum of the mass weighted velocities of all particles/cells in this group, of all types. 
\\
 SubhaloVelDisp
& N
& $ {\rm km/s} $
& One-dimensional velocity dispersion of all the member particles/cells (the 3D dispersion divided by $\sqrt{3}$).
\\
 SubhaloVmax
& N
& $ {\rm km/s} $
& Maximum value of the spherically-averaged rotation curve.
\\
 SubhaloVmaxRad
& N
& $ {\rm kpc}/h $
& Comoving radius of rotation curve maximum (where $V_{max}$ is achieved).
\\
 SubhaloWindMass
& N
& $ 10^{10} {\rm M}_\odot/h $
& Sum of masses of all wind-phase cells in this subhalo (with Type==4 and BirthTime$<=0$). \\ \hline
    \end{tabular}
  \end{center}
\end{table*}

\begin{table*}[htb!]
\footnotesize
  \caption{Description of all fields in the Header group of the group catalog files. Each header field is an attribute.}
  \label{table_gc_header}
  \begin{center}
\renewcommand{\arraystretch}{1.5}
    \begin{tabular}{llp{12cm}}
    \hline
Field & Type & Description \\ \hline\hline
SimulationName
& string
& e.g. 'Illustris-1' or 'Illustris-2-Dark'
\\
 SnapshotNumber
& int
& snapshot number (should be consistent with filename)
\\
 Ngroups\_ThisFile
& int
& Number of groups within this file chunk.
\\
 Nsubgroups\_ThisFile
& int
& Number of subgroups within this file chunk.
\\
 Ngroups\_Total
& int
& Total number of groups for this snapshot.
\\
 Nsubgroups\_Total
& int
& Total number of subgroups for this snapshot.
\\
 NumFiles
& int
& Total number of file chunks the group catalog is split between.
\\
 Num\_ThisFile
& int
& Index of this file chunk (should be consistent with the filename).
\\
 Time
& float
& Scale factor of the snapshot corresponding to this group catalog.
\\
 Redshift
& float
& Redshift of the snapshot corresponding to this group catalog.
\\
 BoxSize
& float
& Side-length of the periodic volume in code units.
\\
 FileOffsets\_Snap
& [$N_{\rm c},6$] int array
& The offset table (by type) for the snapshot files, giving the first particle index in each snap file chunk. Determines which files(s) a given offset+length will cover. 
A two-dimensional array, where the element $(i,j)$ equals the cumulative sum (i.e. offset) of particles of type $i$ in all snapshot file chunks prior to $j$.
\\
 FileOffsets\_Group
& [$N_{\rm c}$] int array
& The offset table for groups in the group catalog files. A one-dimensional array, where the $i^{th}$ element equals the first group number in the $i^{th}$ groupcat file chunk.
\\
 FileOffsets\_Subhalo
& [$N_{\rm c}$] int array
& The offset table for subhalos in the group catalog files. A one-dimensional array, where the $i^{th}$ element equals the first subgroup number in the $i^{th}$ groupcat file chunk.
\\
 FileOffsets\_SubLink
& [$N_{\rm c}$] int array
& The offset table for trees in the {\sc SubLink} files. A one-dimensional array, where the $i^{th}$ element equals the first tree number in the $i^{th}$ {\sc SubLink} file chunk. \\ \hline
    \end{tabular}
  \end{center}
\end{table*}

\begin{table*}[htb!]
\footnotesize
  \caption{Description of all fields in the Offsets group of the group catalog files. Note that all three {\sc LHaloTree} or {\sc SubLink} values equal $-1$ if that subhalo is not in the respective merger tree, which can occur if searching at a snapshot prior to $z=0$. For the offsets, $N_{\rm c}$ indicates the number of file chunks (or pieces) over which that data product has been split.}
  \label{table_gc_offsets}
  \begin{center}
\renewcommand{\arraystretch}{1.5}
    \begin{tabular}{llp{10cm}}
    \hline
Field & Dimensions & Description \\ \hline\hline
Group\_SnapByType
& Ngroups\_Total,6
& The offset table for a given group number (by type), into the snapshot files. That is, the global particle index (across all snap file chunks) of the first particle of this group.
A two-dimensional array, where the element $(i,j)$ equals the cumulative sum (i.e. offset) of particles of type $i$ in all groups prior to group number $j$.
\\
 Group\_FuzzByType
& Ngroups\_Total,6
& Offset into the ``outer fuzz'' (at the end of each snapshot file) for this group.
\\
 Subhalo\_SnapByType
& Nsubgroups\_Total,6
& The offset table for a given subhalo number (by type), into the snapshot files. That is, the global particle index (across all snap file chunks) of the first particle of this subhalo.
A two-dimensional array, where the element $(i,j)$ equals the cumulative sum (i.e. offset) of particles of type $i$ in all subhalos prior to subhalo number $j$.
\\
 Subhalo\_LHaloTreeFile
& Nsubgroups\_Total
& The {\sc LHaloTree} file number with the tree which contains this subhalo.
\\
 Subhalo\_LHaloTreeNum
& Nsubgroups\_Total
& The number of the tree within the above file within which this subhalo is located (e.g. TreeX).
\\
 Subhalo\_LHaloTreeIndex
& Nsubgroups\_Total
& The {\sc LHaloTree} index within the above tree dataset at which this subhalo is located.
\\
 Subhalo\_SublinkRowNum
& Nsubgroups\_Total
& The {\sc SubLink} global index of the location of this subhalo.
\\
 Subhalo\_SublinkSubhaloID
& Nsubgroups\_Total
& The {\sc SubLink} ID of this subhalo.
\\
 Subhalo\_SublinkLastProgenitorID
& Nsubgroups\_Total
& The {\sc SubLink} ID of the last progenitor of this tree (all the subhalos contained in the tree rooted in this subhalo are the ones with IDs between SubhaloID and LastProgenitorID). \\ \hline
    \end{tabular}
  \end{center}
\end{table*}

\begin{table*}[htb!]
\footnotesize
  \caption{Listing of all fields and their descriptions for the {\sc SubLink} merger trees. Note that in addition to the tree fields, all subhalo fields are also 
present, copied exactly from the {\sc Subfind} catalogs. The advantage is that they are ordered in the same order as the tree structure. See the 
group catalog description for their units and descriptions. The Group\_M\_Crit200, Group\_M\_Mean200, and Group\_M\_Tophat200 fields are also present, but are FoF group quantities, such that all subhalos in the same FOF group will have the same value for these three fields.}
  \label{table_sublink}
  \begin{center}
\renewcommand{\arraystretch}{1.5}
    \begin{tabular}{llp{11cm}}
    \hline
Field & Type & Description \\ \hline\hline
SubhaloID
& int64
& Unique identifier of this subhalo, assigned in a ``depth-first'' fashion \citep{lemson06}. This value is contiguous within a single tree.
\\
 SubhaloIDRaw
& int64
& Unique identifier of this subhalo in raw format (= SnapNum$\times 10^{12}$ + SubfindID).
\\
 LastProgenitorID 
& int64
& The SubhaloID of the last progenitor of the tree rooted at this subhalo. Since the SubhaloIDs are assigned in a ``depth-first'' fashion, all the subhalos contained in the tree rooted at this subhalo are the ones with SubhaloIDs between (and including) the SubhaloID and LastProgenitorID of this subhalo. For subhalos with no progenitors, LastProgenitorID == SubhaloID.
\\
 MainLeafProgenitorID 
& int64 
& The SubhaloID of the last progenitor along the main branch, i.e. the earliest progenitor obtained by following the FirstProgenitorID pointer. For subhalos with no progenitors, MainLeafProgenitorID == SubhaloID.
\\
 RootDescendantID 
& int64 
& The SubhaloID of the latest subhalo that can be reached by following the DescendantID link, i.e. the root of the tree to which this subhalo belongs. For subhalos with no descendants, RootDescendantID == SubhaloID.
\\
 TreeID 
& int64 
& Unique identifier of the tree to which this subhalo belongs.
\\
 SnapNum 
& int16 
& The snapshot in which this subhalo is found.
\\
 FirstProgenitorID 
& int64 
& The SubhaloID of this subhalo's first progenitor. The first progenitor is the one with the ``most massive history'' behind it. For subhalos with no progenitors, FirstProgenitorID == -1.
\\
 NextProgenitorID 
& int64 
& The SubhaloID of the subhalo with the next most massive history which shares the same descendant as this subhalo. If there are no more subhalos sharing the same descendant, NextProgenitorID == -1.
\\
 DescendantID 
& int64 
& The SubhaloID of this subhalo's descendant. If this subhalo has no descendants, DescendantID == -1.
\\
 FirstSubhaloInFOFGroupID 
& int64 
& The SubhaloID of the first subhalo (i.e., the one with the most massive history) from the same FOF group.
\\
 NextSubhaloInFOFGroupID 
& int64 
& The SubhaloID of the next subhalo (ordered by their mass history) from the same FOF group. If there are no more subhalos in the same FOF group, NextSubhaloInFOFGroupID == -1.
\\
 NumParticles 
& uint32 
& Number of particles in the current subhalo which were used in the merger tree to determine descendants (e.g. DM-only or stars + star-forming gas).
\\
 Mass 
& float32 
& Mass of the current subhalo, including only the particles which were used in the merger tree to determine descendants (e.g. DM-only or stars + star-forming gas), in units of $ 10^{10} {\rm M}_\odot/h $.
\\
 MassHistory 
& float32 
& Sum of the Mass field of all progenitors along the main branch \citep{delucia07}, in units of $ 10^{10} {\rm M}_\odot/h $.
\\
 SubfindID 
& int32 
& Index of this subhalo in the {\sc Subfind} group catalog. \\ \hline
    \end{tabular}
  \end{center}
\end{table*}

\begin{table*}[htb!]
\footnotesize
  \caption{Listing of all fields in the {\sc LHaloTree} merger trees. Note that in addition to the tree fields, the majority of subhalo fields are also 
present, copied exactly from the {\sc Subfind} catalogs. The advantage is that they are ordered in the same order as the tree structure. See the 
group catalog description for their units and descriptions. The Group\_M\_Crit200, Group\_M\_Mean200, and Group\_M\_Tophat200 fields are also present, but since they are FoF group quantities, all subhalos from the same FOF group will have the same value for these three fields.}
  \label{table_lhalotree}
  \begin{center}
\renewcommand{\arraystretch}{1.5}
    \begin{tabular}{llp{11cm}}
    \hline
Field & Dimensions & Description \\ \hline\hline
\multicolumn{3}{c}{\textbf{Header Groups}}
\\
\hline Redshifts
& \{N\_snap\}
& List of redshifts of the snapshots used to create this merger tree.
\\
 TotNsubhalos
& \{N\_snap\}
& Equal to the number of {\sc Subfind} groups in the group catalog, for each snapshot used to create this merger tree.
\\
 TreeNHalos
& \{N\_halos\}
& The size of \{N\} for each TreeX group in this file, e.g. the total number of halos (across time) in that group.
\\
 FirstSnapshotNr
& 1
& First snapshot number used to make these merger trees (should be 0).
\\
 LastSnapshotNr
& 1
& Last snapshot number used to make these merger trees (should be 135).
\\
 SnapSkipFac
& 1
& Snapshot stride when making these merger trees (should be 1).
\\
 NtreesPerFile
& 1
& The size of \{N\_halos\} for this file, can be used to calculate the offset to map a FoF group number to a TreeX group name (made to be roughly equal across chunks).
\\
 NhalosPerFile
& 1
& The total number of tree members (subhalos) in this file. Equals the sum of all elements of TreeNHalos.
\\
 ParticleMass
& 1
& The dark matter particle mass used to make these merger trees, in units of $10^{10} {\rm M}_\odot / h$.
\\ 
\hline
\multicolumn{3}{c}{\textbf{TreeX Groups}}
\\ 
\hline\rule{0pt}{3ex}SubhaloNumber 
& (N) 
& The ID of this subhalo, unique within the full simulation for this snapshot. Indexes the {\sc Subfind} group catalog at SnapNum.
\\
 Descendant 
& (N) 
& The index of the subhalo's descendant within the merger tree, if any (-1 otherwise). Indexes this TreeX group.
\\
 FirstProgenitor 
& (N) 
& The index of the subhalo's first progenitor within the merger tree, if any (-1 otherwise). The first progenitor is defined as the most massive one. (-1 if none) Indexes this TreeX group.
\\
 NextProgenitor 
& (N) 
& The index of the next subhalo from the same snapshot which shares the same descendant, if any (-1 if this is the last). Indexes this TreeX group.
\\
 FirstHaloInFOFGroup 
& (N) 
& The index of the main subhalo (i.e. the most massive one) from the same FOF group.  Indexes this TreeX group.
\\
 NextHaloInFOFGroup 
& (N) 
& The index of the next subhalo from the same FOF group (-1 if this is the last). Indexes this TreeX group.
\\
 FileNr 
& (N) 
& File number in which the subhalo is found. Redundant, i.e. for a given [chunkNum] file, this array will be constant and equal to [chunkNum].
\\
 SnapNum 
& (N) 
& The snapshot in which this subhalo was found. \\ \hline
    \end{tabular}
  \end{center}
\end{table*}

\clearpage
\section*{Appendix C: Supplementary Data Details}

\setcounter{table}{0}
\renewcommand*{\theHtable}{C.\arabic{table}}
\gdef\thetable{C.\arabic{table}}

\begin{table*}[htb!]
\footnotesize
  \caption{Details of the supplementary data catalog: Photometric Non-Parametric Stellar Morphologies. The four bands which replace {band\_name} are: gSDSS, iSDSS, uSDSS, and hWFC3 (WFC3-IR/F160W). The four camera views are indexed 0, 1, 2, and 3.}
  \label{table_supp_nonpara_morphs}
  \begin{center}
\renewcommand{\arraystretch}{1.5}
    \begin{tabular}{lcp{10cm}}
    \hline
Group Name & Units & Description \\ \hline\hline
/Snapshot\_135/SubfindID\_cam{0,1,2,3}
& -
& The {\sc Subfind} IDs these values correspond to (different for each camera view, but the same for all bands and fields). {10654,10618,10639,10620} entries.
\\
 /Snapshot\_135/{band\_name}/Gini\_cam{0,1,2,3}
& -
& The $G\_{\rm ini}$ coefficient, which measures the relative distribution of the galaxy pixel flux values.
\\
 /Snapshot\_135/{band\_name}/M20\_cam{0,1,2,3}
& -
& $M_{20}$, the second-order moment of the brightest 20\% of the galaxy's flux.
\\
 /Snapshot\_135/{band\_name}/C\_cam{0,1,2,3}
& -
& The concentration parameter $C$.
\\
 /Snapshot\_135/{band\_name}/RP\_cam{0,1,2,3}
& $kpc$
& The elliptical Petrosian radius $r_P$.
\\
 /Snapshot\_135/{band\_name}/RE\_cam{0,1,2,3}
& $kpc$
& The elliptical half-light radius $r_E$.
\\ \hline
    \end{tabular}
  \end{center}
\end{table*}

\begin{table*}[htb!]
\footnotesize
  \caption{Details of the supplementary data catalog: Stellar Circularities, Angular Momenta, and Axis Ratios. Note that, in addition to these values which are measured within 10$R_E$, several fields are also computed including all stars in the subhalo, and are available as the ``\_allstars'' datasets.}
  \label{table_supp_stellar_circs}
  \begin{center}
\renewcommand{\arraystretch}{1.5}
    \begin{tabular}{p{3cm}cp{12cm}}
    \hline
Group Name & Units & Description \\ \hline\hline
/Snapshot\_N/ \newline SubfindID
& -
& The {\sc Subfind} IDs these values correspond to (27345 entries).
\\ 
\rule{0pt}{3ex}/Snapshot\_N/ \newline SpecificAngMom
& $ {\rm km/s} \times {\rm kpc} $
& The specific angular momentum of the stars.
\\
\rule{0pt}{3ex}/Snapshot\_N/ \newline CircAbove07Frac
& -
& The fraction of stars with $ \epsilon > 0.7 $. This is a common definition of the disk stars - those with significant (positive) rotational support.
\\
\rule{0pt}{3ex}/Snapshot\_N/ \newline  CircAbove07\newline MinusBelowNeg07Frac
& -
& The fraction of stars with $ \epsilon > 0.7 $ minus the fraction of stars with $ \epsilon < -0.7 $. This removes the contribution of the bulge to the disk, assuming the bulge is symmetric around $ \epsilon=0 $.
\\
\rule{0pt}{3ex}/Snapshot\_N/ \newline CircTwiceBelow0Frac
& -
&  The fraction of stars with $ \epsilon < 0 $, multiplied by two. This is another common way in the literature to define the bulge.
\\
\rule{0pt}{3ex}/Snapshot\_N/ \newline MassTensorEigenVals
& $ {\rm kpc} $
& Three numbers for each galaxy, which are the eigenvalues of the mass tensor of the stellar mass inside the stellar $2R_{1/2}$. This means that in a coordinate system that is aligned with the eigenvectors (principal axes), the component $i$ equals $M_i\equiv\sqrt{\sum\limits_j m_jr_{j,i}^2}/\sqrt{\sum\limits_j m_j}$, where $j$ enumerates over stellar particles inside that radius, $r_{j,i}$ is the distance of stellar particle $j$ in the $i$ axis from the most bound particle of the galaxy, and $m_j$ is its mass, and $i\in(1,2,3)$. They are sorted such that $M_1<M_2<M_3$. Example use: $M_1/\sqrt{M_2M_3}$ can represent the flatness of the galaxy.
\\
\rule{0pt}{3ex}/Snapshot\_N/ \newline ReducedMass\newline TensorEigenVals
& -
& Similar to the above, except less weight is given to further away particles. The orientation of the system is the same, but the quantity measured for each axis is instead $M_i\equiv\sqrt{\sum\limits_j m_jr_{j,i}^2/R_j^2}/\sqrt{\sum\limits_j m_j}$, where $R_j\equiv\sum\limits_i r_{j,i}^2$ is the distance of star $j$ from the centre of the galaxy. \\ \hline
    \end{tabular}
  \end{center}
\end{table*}

\onecolumn
\clearpage
\section*{Appendix D: API Examples and Reference}

\setcounter{table}{0}
\renewcommand*{\theHtable}{D.\arabic{table}}
\gdef\thetable{D.\arabic{table}}

To be explicit by way of example, the following are absolute URLs for the Illustris API covering some of its functionality, 
where the type of the request should be clear from the preceding documentation.

\begin{itemize}[leftmargin=2em]
\footnotesize
\setlength\itemsep{0em}
\item \url{http://www.illustris-project.org/api/Illustris-2/}
\item \url{http://www.illustris-project.org/api/Illustris-2/snapshots/68/}
\item \url{http://www.illustris-project.org/api/Illustris-1/snapshots/135/subhalos/73664/}
\item \url{http://www.illustris-project.org/api/Illustris-1/snapshots/135/subhalos/73664/stellar_mocks/broadband.fits}
\item \url{http://www.illustris-project.org/api/Illustris-1/snapshots/135/subhalos/73664/stellar_mocks/sed.txt}
\item \url{http://www.illustris-project.org/api/Illustris-1/snapshots/80/halos/523312/cutout.hdf5?dm=Coordinates&gas=all}
\item \url{http://www.illustris-project.org/api/Illustris-3/snapshots/135/subhalos?mass__gt=10.0&mass__lt=20.0}
\item \url{http://www.illustris-project.org/api/Illustris-2/snapshots/68/subhalos/50000/sublink/full.hdf5}
\item \url{http://www.illustris-project.org/api/Illustris-2/snapshots/68/subhalos/50000/sublink/mpb.json}
\item \url{http://www.illustris-project.org/api/Illustris-1/files/groupcat-135.5.hdf5}
\item \url{http://www.illustris-project.org/api/Illustris-2/files/snapshot-135.10.hdf5}
\item \url{http://www.illustris-project.org/api/Illustris-2/files/snapshot-135.10.hdf5?dm=all}
\item \url{http://www.illustris-project.org/api/Illustris-3/files/sublink.2.hdf5}
\end{itemize}

In the online documentation we provide a complete getting started guide for the web-based API, as well as a cookbook of common tasks, in Python, IDL, and Matlab. 
Here we include just four examples taken from that documentation, and only in Python, to give a flavor of the approach. The task numbers are taken from the online version.

% MANUAL PYGMENTS VERSION:
% ------------------------
\vspace{7mm}
\noindent\textbf{Task 0:} First, we define a helper function, to make the HTTP response, and check for errors. If the response is JSON, automatically parse it. If the 
response is binary data, automatically save it to a file.
\begin{Verbatim}[commandchars=\\\{\}]
\PY{o}{\PYZgt{}\PYZgt{}}\PY{o}{\PYZgt{}} \PY{k}{def} \PY{n+nf}{get}\PY{p}{(}\PY{n}{path}\PY{p}{,} \PY{n}{params}\PY{o}{=}\PY{n+nb+bp}{None}\PY{p}{)}\PY{p}{:}
\PY{o}{\PYZgt{}\PYZgt{}}\PY{o}{\PYZgt{}}     \PY{c}{\PYZsh{} make HTTP GET request to path}
\PY{o}{\PYZgt{}\PYZgt{}}\PY{o}{\PYZgt{}}     \PY{n}{headers} \PY{o}{=} \PY{p}{\PYZob{}}\PY{l+s}{\PYZdq{}}\PY{l+s}{api\PYZhy{}key}\PY{l+s}{\PYZdq{}}\PY{p}{:}\PY{l+s}{\PYZdq{}}\PY{l+s}{INSERT\PYZus{}API\PYZus{}KEY\PYZus{}HERE}\PY{l+s}{\PYZdq{}}\PY{p}{\PYZcb{}}
\PY{o}{\PYZgt{}\PYZgt{}}\PY{o}{\PYZgt{}}     \PY{n}{r} \PY{o}{=} \PY{n}{requests}\PY{o}{.}\PY{n}{get}\PY{p}{(}\PY{n}{path}\PY{p}{,} \PY{n}{params}\PY{o}{=}\PY{n}{params}\PY{p}{,} \PY{n}{headers}\PY{o}{=}\PY{n}{headers}\PY{p}{)}
\PY{o}{\PYZgt{}\PYZgt{}}\PY{o}{\PYZgt{}}
\PY{o}{\PYZgt{}\PYZgt{}}\PY{o}{\PYZgt{}}     \PY{c}{\PYZsh{} raise exception if response code is not HTTP SUCCESS (200)}
\PY{o}{\PYZgt{}\PYZgt{}}\PY{o}{\PYZgt{}}     \PY{n}{r}\PY{o}{.}\PY{n}{raise\PYZus{}for\PYZus{}status}\PY{p}{(}\PY{p}{)}
\PY{o}{\PYZgt{}\PYZgt{}}\PY{o}{\PYZgt{}}
\PY{o}{\PYZgt{}\PYZgt{}}\PY{o}{\PYZgt{}}     \PY{k}{if} \PY{n}{r}\PY{o}{.}\PY{n}{headers}\PY{p}{[}\PY{l+s}{\PYZsq{}}\PY{l+s}{content\PYZhy{}type}\PY{l+s}{\PYZsq{}}\PY{p}{]} \PY{o}{==} \PY{l+s}{\PYZsq{}}\PY{l+s}{application/json}\PY{l+s}{\PYZsq{}}\PY{p}{:}
\PY{o}{\PYZgt{}\PYZgt{}}\PY{o}{\PYZgt{}}         \PY{k}{return} \PY{n}{r}\PY{o}{.}\PY{n}{json}\PY{p}{(}\PY{p}{)} \PY{c}{\PYZsh{} parse json responses automatically}
\PY{o}{\PYZgt{}\PYZgt{}}\PY{o}{\PYZgt{}}
\PY{o}{\PYZgt{}\PYZgt{}}\PY{o}{\PYZgt{}}     \PY{k}{if} \PY{l+s}{\PYZsq{}}\PY{l+s}{content\PYZhy{}disposition}\PY{l+s}{\PYZsq{}} \PY{o+ow}{in} \PY{n}{r}\PY{o}{.}\PY{n}{headers}\PY{p}{:}
\PY{o}{\PYZgt{}\PYZgt{}}\PY{o}{\PYZgt{}}         \PY{n}{filename} \PY{o}{=} \PY{n}{r}\PY{o}{.}\PY{n}{headers}\PY{p}{[}\PY{l+s}{\PYZsq{}}\PY{l+s}{content\PYZhy{}disposition}\PY{l+s}{\PYZsq{}}\PY{p}{]}\PY{o}{.}\PY{n}{split}\PY{p}{(}\PY{l+s}{\PYZdq{}}\PY{l+s}{filename=}\PY{l+s}{\PYZdq{}}\PY{p}{)}\PY{p}{[}\PY{l+m+mi}{1}\PY{p}{]}
\PY{o}{\PYZgt{}\PYZgt{}}\PY{o}{\PYZgt{}}         \PY{k}{with} \PY{n+nb}{open}\PY{p}{(}\PY{n}{filename}\PY{p}{,} \PY{l+s}{\PYZsq{}}\PY{l+s}{wb}\PY{l+s}{\PYZsq{}}\PY{p}{)} \PY{k}{as} \PY{n}{f}\PY{p}{:}
\PY{o}{\PYZgt{}\PYZgt{}}\PY{o}{\PYZgt{}}             \PY{n}{f}\PY{o}{.}\PY{n}{write}\PY{p}{(}\PY{n}{r}\PY{o}{.}\PY{n}{content}\PY{p}{)}
\PY{o}{\PYZgt{}\PYZgt{}}\PY{o}{\PYZgt{}}         \PY{k}{return} \PY{n}{filename} \PY{c}{\PYZsh{} return the filename string}
\end{Verbatim}

\vspace{5mm}\noindent\textbf{Task 1:} For Illustris-1 at $z=0$, get all the fields available for the subhalo with id=0 and print its total mass and stellar half mass radius.
\begin{Verbatim}[commandchars=\\\{\}]
\PY{o}{\PYZgt{}\PYZgt{}}\PY{o}{\PYZgt{}} \PY{n}{url} \PY{o}{=} \PY{l+s}{\PYZdq{}}\PY{l+s}{http://www.illustris\PYZhy{}project.org/api/Illustris\PYZhy{}1/snapshots/135/subhalos/0/}\PY{l+s}{\PYZdq{}}
\PY{o}{\PYZgt{}\PYZgt{}}\PY{o}{\PYZgt{}} \PY{n}{r} \PY{o}{=} \PY{n}{get}\PY{p}{(}\PY{n}{url}\PY{p}{)}
\PY{o}{\PYZgt{}\PYZgt{}}\PY{o}{\PYZgt{}} \PY{n}{r}\PY{p}{[}\PY{l+s}{\PYZsq{}}\PY{l+s}{mass}\PY{l+s}{\PYZsq{}}\PY{p}{]}
\PY{l+m+mf}{22174.8}

\PY{o}{\PYZgt{}\PYZgt{}}\PY{o}{\PYZgt{}} \PY{n}{r}\PY{p}{[}\PY{l+s}{\PYZsq{}}\PY{l+s}{halfmassrad\PYZus{}stars}\PY{l+s}{\PYZsq{}}\PY{p}{]}
\PY{l+m+mf}{12.395}
\end{Verbatim}

\vspace{10mm}
\noindent\textbf{Task 2:} For Illustris-1 at $z=2$, search for all subhalos with total mass $10^{11.9} {\rm M}_\odot < M < 10^{12.1} {\rm M}_\odot$, print the number returned, and the {\sc Subfind} IDs of the first five results.
\begin{Verbatim}[commandchars=\\\{\}]
\PY{o}{\PYZgt{}\PYZgt{}}\PY{o}{\PYZgt{}} \PY{c}{\PYZsh{} first convert log solar masses into group catalog units}
\PY{o}{\PYZgt{}\PYZgt{}}\PY{o}{\PYZgt{}} \PY{n}{mass\PYZus{}min} \PY{o}{=} \PY{l+m+mi}{10}\PY{o}{*}\PY{o}{*}\PY{l+m+mf}{11.9} \PY{o}{/} \PY{l+m+mf}{1e10} \PY{o}{*} \PY{l+m+mf}{0.704}
\PY{o}{\PYZgt{}\PYZgt{}}\PY{o}{\PYZgt{}} \PY{n}{mass\PYZus{}max} \PY{o}{=} \PY{l+m+mi}{10}\PY{o}{*}\PY{o}{*}\PY{l+m+mf}{12.1} \PY{o}{/} \PY{l+m+mf}{1e10} \PY{o}{*} \PY{l+m+mf}{0.704}
\PY{o}{\PYZgt{}\PYZgt{}}\PY{o}{\PYZgt{}} 
\PY{o}{\PYZgt{}\PYZgt{}}\PY{o}{\PYZgt{}} \PY{n}{params} \PY{o}{=} \PY{p}{\PYZob{}}\PY{l+s}{\PYZsq{}}\PY{l+s}{mass\PYZus{}\PYZus{}gt}\PY{l+s}{\PYZsq{}}\PY{p}{:}\PY{n}{mass\PYZus{}min}\PY{p}{,} \PY{l+s}{\PYZsq{}}\PY{l+s}{mass\PYZus{}\PYZus{}lt}\PY{l+s}{\PYZsq{}}\PY{p}{:}\PY{n}{mass\PYZus{}max}\PY{p}{\PYZcb{}}
\PY{o}{\PYZgt{}\PYZgt{}}\PY{o}{\PYZgt{}} 
\PY{o}{\PYZgt{}\PYZgt{}}\PY{o}{\PYZgt{}} \PY{c}{\PYZsh{} make the request}
\PY{o}{\PYZgt{}\PYZgt{}}\PY{o}{\PYZgt{}} \PY{n}{url} \PY{o}{=} \PY{l+s}{\PYZdq{}}\PY{l+s}{http://www.illustris\PYZhy{}project.org/api/Illustris\PYZhy{}1/snapshots/z=2/subhalos/}\PY{l+s}{\PYZdq{}}
\PY{o}{\PYZgt{}\PYZgt{}}\PY{o}{\PYZgt{}} \PY{n}{subhalos} \PY{o}{=} \PY{n}{get}\PY{p}{(}\PY{n}{url}\PY{p}{,} \PY{n}{params}\PY{p}{)}
\PY{o}{\PYZgt{}\PYZgt{}}\PY{o}{\PYZgt{}} \PY{n}{subhalos}\PY{p}{[}\PY{l+s}{\PYZsq{}}\PY{l+s}{count}\PY{l+s}{\PYZsq{}}\PY{p}{]}
\PY{l+m+mi}{550}

\PY{o}{\PYZgt{}\PYZgt{}}\PY{o}{\PYZgt{}} \PY{n}{ids} \PY{o}{=} \PY{p}{[} \PY{n}{subhalos}\PY{p}{[}\PY{l+s}{\PYZsq{}}\PY{l+s}{results}\PY{l+s}{\PYZsq{}}\PY{p}{]}\PY{p}{[}\PY{n}{i}\PY{p}{]}\PY{p}{[}\PY{l+s}{\PYZsq{}}\PY{l+s}{id}\PY{l+s}{\PYZsq{}}\PY{p}{]} \PY{k}{for} \PY{n}{i} \PY{o+ow}{in} \PY{n+nb}{range}\PY{p}{(}\PY{l+m+mi}{5}\PY{p}{)} \PY{p}{]}
\PY{o}{\PYZgt{}\PYZgt{}}\PY{o}{\PYZgt{}} \PY{n}{ids}
\PY{p}{[}\PY{l+m+mi}{1}\PY{p}{,} \PY{l+m+mi}{1352}\PY{p}{,} \PY{l+m+mi}{5525}\PY{p}{,} \PY{l+m+mi}{6574}\PY{p}{,} \PY{l+m+mi}{12718}\PY{p}{]}
\end{Verbatim}

\vspace{5mm}
\noindent\textbf{Task 8:} For Illustris-1 at $z=2$, for five specific {\sc Subfind} IDs (from above: 1, 1352, 5525, 6574, 12718), locate the $z=0$ descendant of each by using the API to walk down the {\sc SubLink} descendant links.
\begin{Verbatim}[commandchars=\\\{\}]
\PY{o}{\PYZgt{}\PYZgt{}}\PY{o}{\PYZgt{}} \PY{n}{ids} \PY{o}{=} \PY{p}{[}\PY{l+m+mi}{1}\PY{p}{,} \PY{l+m+mi}{1352}\PY{p}{,} \PY{l+m+mi}{5525}\PY{p}{,} \PY{l+m+mi}{6574}\PY{p}{,} \PY{l+m+mi}{12718}\PY{p}{]}
\PY{o}{\PYZgt{}\PYZgt{}}\PY{o}{\PYZgt{}} \PY{n}{z0\PYZus{}descendant\PYZus{}ids} \PY{o}{=} \PY{p}{[}\PY{o}{\PYZhy{}}\PY{l+m+mi}{1}\PY{p}{]}\PY{o}{*}\PY{n+nb}{len}\PY{p}{(}\PY{n}{ids}\PY{p}{)}
\PY{o}{\PYZgt{}\PYZgt{}}\PY{o}{\PYZgt{}} 
\PY{o}{\PYZgt{}\PYZgt{}}\PY{o}{\PYZgt{}} \PY{k}{for} \PY{n}{i}\PY{p}{,}\PY{n+nb}{id} \PY{o+ow}{in} \PY{n+nb}{enumerate}\PY{p}{(}\PY{n}{ids}\PY{p}{)}\PY{p}{:}
\PY{o}{\PYZgt{}\PYZgt{}}\PY{o}{\PYZgt{}}     \PY{n}{start\PYZus{}url} \PY{o}{=} \PY{l+s}{\PYZdq{}}\PY{l+s}{http://www.illustris\PYZhy{}project.org/api/Illustris\PYZhy{}1/snapshots/z=2/subhalos/}\PY{l+s}{\PYZdq{}}
\PY{o}{\PYZgt{}\PYZgt{}}\PY{o}{\PYZgt{}}     \PY{n}{start\PYZus{}url} \PY{o}{+}\PY{o}{=} \PY{n+nb}{str}\PY{p}{(}\PY{n+nb}{id}\PY{p}{)}
\PY{o}{\PYZgt{}\PYZgt{}}\PY{o}{\PYZgt{}}     \PY{n}{sub} \PY{o}{=} \PY{n}{get}\PY{p}{(}\PY{n}{start\PYZus{}url}\PY{p}{)}
\PY{o}{\PYZgt{}\PYZgt{}}\PY{o}{\PYZgt{}}     
\PY{o}{\PYZgt{}\PYZgt{}}\PY{o}{\PYZgt{}}     \PY{k}{while} \PY{n}{sub}\PY{p}{[}\PY{l+s}{\PYZsq{}}\PY{l+s}{desc\PYZus{}sfid}\PY{l+s}{\PYZsq{}}\PY{p}{]} \PY{o}{!=} \PY{o}{\PYZhy{}}\PY{l+m+mi}{1}\PY{p}{:}
\PY{o}{\PYZgt{}\PYZgt{}}\PY{o}{\PYZgt{}}         \PY{c}{\PYZsh{} request the full subhalo details of the descendant by following the sublink URL}
\PY{o}{\PYZgt{}\PYZgt{}}\PY{o}{\PYZgt{}}         \PY{n}{sub} \PY{o}{=} \PY{n}{get}\PY{p}{(}\PY{n}{sub}\PY{p}{[}\PY{l+s}{\PYZsq{}}\PY{l+s}{related}\PY{l+s}{\PYZsq{}}\PY{p}{]}\PY{p}{[}\PY{l+s}{\PYZsq{}}\PY{l+s}{sublink\PYZus{}descendant}\PY{l+s}{\PYZsq{}}\PY{p}{]}\PY{p}{)}
\PY{o}{\PYZgt{}\PYZgt{}}\PY{o}{\PYZgt{}}         \PY{k}{if} \PY{n}{sub}\PY{p}{[}\PY{l+s}{\PYZsq{}}\PY{l+s}{snap}\PY{l+s}{\PYZsq{}}\PY{p}{]} \PY{o}{==} \PY{l+m+mi}{135}\PY{p}{:}
\PY{o}{\PYZgt{}\PYZgt{}}\PY{o}{\PYZgt{}}             \PY{n}{z0\PYZus{}descendant\PYZus{}ids}\PY{p}{[}\PY{n}{i}\PY{p}{]} \PY{o}{=} \PY{n}{sub}\PY{p}{[}\PY{l+s}{\PYZsq{}}\PY{l+s}{id}\PY{l+s}{\PYZsq{}}\PY{p}{]}
\PY{o}{\PYZgt{}\PYZgt{}}\PY{o}{\PYZgt{}}             
\PY{o}{\PYZgt{}\PYZgt{}}\PY{o}{\PYZgt{}}     \PY{k}{if} \PY{n}{z0\PYZus{}descendant\PYZus{}ids}\PY{p}{[}\PY{n}{i}\PY{p}{]} \PY{o}{\PYZgt{}}\PY{o}{=} \PY{l+m+mi}{0}\PY{p}{:} \PY{c}{\PYZsh{} note: possible that descendant branch did not reach z=0}
\PY{o}{\PYZgt{}\PYZgt{}}\PY{o}{\PYZgt{}}         \PY{k}{print} \PY{l+s}{\PYZsq{}}\PY{l+s}{Descendant of }\PY{l+s}{\PYZsq{}} \PY{o}{+} \PY{n+nb}{str}\PY{p}{(}\PY{n+nb}{id}\PY{p}{)} \PY{o}{+} \PY{l+s}{\PYZsq{}}\PY{l+s}{ at z=0 is }\PY{l+s}{\PYZsq{}} \PY{o}{+} \PY{n+nb}{str}\PY{p}{(}\PY{n}{z0\PYZus{}descendant\PYZus{}ids}\PY{p}{[}\PY{n}{i}\PY{p}{]}\PY{p}{)}

\PY{n}{Descendant} \PY{n}{of} \PY{l+m+mi}{1} \PY{n}{at} \PY{n}{z}\PY{o}{=}\PY{l+m+mi}{0} \PY{o+ow}{is} \PY{l+m+mi}{30465}
\PY{n}{Descendant} \PY{n}{of} \PY{l+m+mi}{1352} \PY{n}{at} \PY{n}{z}\PY{o}{=}\PY{l+m+mi}{0} \PY{o+ow}{is} \PY{l+m+mi}{41396}
\PY{n}{Descendant} \PY{n}{of} \PY{l+m+mi}{5525} \PY{n}{at} \PY{n}{z}\PY{o}{=}\PY{l+m+mi}{0} \PY{o+ow}{is} \PY{l+m+mi}{99148}
\PY{n}{Descendant} \PY{n}{of} \PY{l+m+mi}{6574} \PY{n}{at} \PY{n}{z}\PY{o}{=}\PY{l+m+mi}{0} \PY{o+ow}{is} \PY{l+m+mi}{51811}
\PY{n}{Descendant} \PY{n}{of} \PY{l+m+mi}{12718} \PY{n}{at} \PY{n}{z}\PY{o}{=}\PY{l+m+mi}{0} \PY{o+ow}{is} \PY{l+m+mi}{194303}
\end{Verbatim}

\vspace{5mm}
\noindent\textbf{Task 11:} Download the entire Illustris-1 $z=0$ snapshot including \textit{only the positions, masses, and metallicities of stars} (in the form of 512 HDF5 files). 
In this example, since we only need these three fields for stars only, we can reduce the download and storage size from $\sim$1.5\,TB to $\sim$17\,GB.
\begin{Verbatim}[commandchars=\\\{\}]
\PY{o}{\PYZgt{}\PYZgt{}}\PY{o}{\PYZgt{}} \PY{n}{base\PYZus{}url} \PY{o}{=} \PY{l+s}{\PYZdq{}}\PY{l+s}{http://www.illustris\PYZhy{}project.org/api/Illustris\PYZhy{}1/}\PY{l+s}{\PYZdq{}}
\PY{o}{\PYZgt{}\PYZgt{}}\PY{o}{\PYZgt{}} \PY{n}{sim\PYZus{}metadata} \PY{o}{=} \PY{n}{get}\PY{p}{(}\PY{n}{base\PYZus{}url}\PY{p}{)}
\PY{o}{\PYZgt{}\PYZgt{}}\PY{o}{\PYZgt{}} \PY{n}{params} \PY{o}{=} \PY{p}{\PYZob{}}\PY{l+s}{\PYZsq{}}\PY{l+s}{stars}\PY{l+s}{\PYZsq{}}\PY{p}{:}\PY{l+s}{\PYZsq{}}\PY{l+s}{Coordinates,Masses,GFM\PYZus{}Metallicity}\PY{l+s}{\PYZsq{}}\PY{p}{\PYZcb{}}
\PY{o}{\PYZgt{}\PYZgt{}}\PY{o}{\PYZgt{}} 
\PY{o}{\PYZgt{}\PYZgt{}}\PY{o}{\PYZgt{}} \PY{k}{for} \PY{n}{i} \PY{o+ow}{in} \PY{n+nb}{range}\PY{p}{(}\PY{n}{sim\PYZus{}metadata}\PY{p}{[}\PY{l+s}{\PYZsq{}}\PY{l+s}{num\PYZus{}files\PYZus{}snapshot}\PY{l+s}{\PYZsq{}}\PY{p}{]}\PY{p}{)}\PY{p}{:}
\PY{o}{\PYZgt{}\PYZgt{}}\PY{o}{\PYZgt{}}     \PY{n}{file\PYZus{}url} \PY{o}{=} \PY{n}{base\PYZus{}url} \PY{o}{+} \PY{l+s}{\PYZdq{}}\PY{l+s}{files/snapshot\PYZhy{}135.}\PY{l+s}{\PYZdq{}} \PY{o}{+} \PY{n+nb}{str}\PY{p}{(}\PY{n}{i}\PY{p}{)} \PY{o}{+} \PY{l+s}{\PYZdq{}}\PY{l+s}{.hdf5}\PY{l+s}{\PYZdq{}}
\PY{o}{\PYZgt{}\PYZgt{}}\PY{o}{\PYZgt{}}     \PY{n}{saved\PYZus{}filename} \PY{o}{=} \PY{n}{get}\PY{p}{(}\PY{n}{file\PYZus{}url}\PY{p}{,} \PY{n}{params}\PY{p}{)}
\PY{o}{\PYZgt{}\PYZgt{}}\PY{o}{\PYZgt{}}     \PY{k}{print} \PY{n}{saved\PYZus{}filename}
\end{Verbatim}

\clearpage

\begin{table*}[htb!]
\footnotesize
  \caption{API Endpoint Descriptions and Reference (I): simulation and snapshot meta-data, subhalos and halos, merger trees.}
  \label{table_api1}
  \begin{center}
\renewcommand{\arraystretch}{1.5}
    \begin{tabular}{p{3.5cm}p{10cm}l}
    \hline
Endpoint & Description & Return Type \\ \hline\hline
/api/
& list all simulations currently accessible to the user
& json,api (?format=)
\\ 
 /api/\{sim\_name\}/
& list metadata (including list of all snapshots+redshifts) for \{sim\_name\}
& json,api (?format=)
\\ 
 /api/\{sim\_name\}/\newline snapshots/
& list all snapshots which exist for this simulation
& json,api (?format=)
\\ 
 /api/\{sim\_name\}/\newline snapshots/\{num\}/
& list metadata for snapshot \{num\} of simulation \{sim\_name\}
& json,api (?format=)
\\ 
 /api/\{sim\_name\}/\newline snapshots/z=\{redshift\}/
& redirect to the snapshot which exists closest to \{redshift\} (with a maximum allowed error of 0.1 in redshift)
& json,api (?format=)
\\
\hline
\multicolumn{3}{l}{define {[}base{]} = /api/\{sim\_name\}/snapshots/\{num\} or {[}base{]} = /api/\{sim\_name\}/snapshots/z=\{redshift\}} \\
\multicolumn{3}{l}{(after selection of a particular simulation and snapshot)}
\\
\hline
\multicolumn{3}{c}{\textbf{Subfind Subhalos}}
\\ 
\hline{[}base{]}/subhalos/
& paginated list of all subhalos for this snapshot of this run
& json,api (?format=)
\\
 {[}base{]}/subhalos/\newline ?\{search\_query\}
& execute \{search\_query\} over all subhalos, return those satisfying the search with basic fields and links to /subhalos/\{id\}
& json,api (?format=)
\\
 {[}base{]}/subhalos/\newline  \{id\}
& list available data fields and links to all queries possible on {\sc Subfind} subhalo \{id\}
& json,api (?format=)
\\ 
 {[}base{]}/subhalos/\newline  \{id\}/info.json
& extract all group catalog fields for subhalo \{id\}
& json (.ext)
\\ 
 {[}base{]}/subhalos/\newline  \{id\}/cutout.hdf5
& return snapshot cutout of subhalo \{id\}, all particle types and fields
& HDF5 (.ext)
\\ 
 {[}base{]}/subhalos/\newline  \{id\}/cutout.hdf5 \newline ?\{cutout\_query\}
& return snapshot cutout of subhalo \{id\} corresponding to the \{cutout\_query\}
& HDF5 (.ext)
\\ 
\hline
\multicolumn{3}{c}{\textbf{FoF Halos}}
\\ 
\hline\rule{0pt}{3ex}{[}base{]}/halos/\{halo\_id\}/
& list what we know about this FoF halo, in particular the 'child\_subhalos'
& json,api (?format=)
\\ 
 {[}base{]}/halos/\{halo\_id\}/ \newline info.json
& extract all group catalog fields for halo \{halo\_id\}
& json (.ext)
\\ 
 {[}base{]}/halos/\{halo\_id\}/ \newline cutout.hdf5
& return snapshot cutout of halo \{halo\_id\}, all particle types and fields
& HDF5 (.ext)
\\ 
 {[}base{]}/halos/\{halo\_id\}/ \newline cutout.hdf5?\{cutout\_query\}
& return snapshot cutout of halo \{halo\_id\} corresponding to the \{cutout\_query\}
& HDF5 (.ext)
\\
\hline
\multicolumn{3}{c}{\textbf{Merger Trees}}
\\ 
\hline
\rule{0pt}{3ex}{[}base{]}/subhalos/\{id\}/ \newline lhalotree/full.hdf5
& retrieve full tree (flat HDF5 format or hierchical/nested JSON format)
& HDF5,json (.ext)
\\ 
 {[}base{]}/subhalos/\{id\}/ \newline lhalotree/mpb.hdf5
& retrieve only main progenitor branch (towards higher redshift for this subhalo)
& HDF5,json (.ext)
\\ 
 {[}base{]}/subhalos/\{id\}/ \newline sublink/full.hdf5
& same as above for 'lhalotree' but for sublink
& HDF5,json (.ext)
\\ 
 {[}base{]}/subhalos/\{id\}/ \newline sublink/mpb.hdf5
& same as above for 'lhalotree' but for sublink
& HDF5,json (.ext)
\\ \hline
    \end{tabular}
  \end{center}
\end{table*}

\begin{table*}[htb!]
\footnotesize
  \caption{API Endpoint Descriptions and Reference (II): supplementary data catalogs, file downloads.}
  \label{table_api2}
  \begin{center}
\renewcommand{\arraystretch}{1.5}
    \begin{tabular}{p{4cm}p{10cm}l}
    \hline
Endpoint & Description & Return Type \\ \hline
\\
\hline
\multicolumn{3}{c}{\textbf{supplementary data: stellar mocks}}
\\ 
\hline\rule{0pt}{3ex}{[}base{]}/subhalos/\{id\}/ \newline stellar\_mocks/broadband.fits
& download raw broadband fits file for subhalo \{id\}
& FITS (.ext)
\\
 {[}base{]}/subhalos/\{id\}/ \newline stellar\_mocks/broadband.hdf5? \newline view=\{view\}
& download subset of broadband fits file for subhalo \{id\}: all 36 bands for view number \{view\}
& HDF5 (.ext)
\\
 {[}base{]}/subhalos/\{id\}/ \newline stellar\_mocks/broadband.hdf5? \newline band={band}
& download subset of broadband fits file for subhalo \{id\}: all 4 views for band \{band\} (1-indexed number, or name)
& HDF5 (.ext)
\\
 {[}base{]}/subhalos/\{id\}/ \newline stellar\_mocks/image.png
& download stellar mock png 2D image (subhalo particles only)
& PNG (.ext)
\\
 {[}base{]}/subhalos/\{id\}/ \newline stellar\_mocks/image\_fof.png
& download stellar mock png 2D image (all group particles)
& PNG (.ext)
\\
 {[}base{]}/subhalos/\{id\}/ \newline stellar\_mocks/image\_gz.png
& download stellar mock png 2D image ('galaxy zoo' image w/ realistic noise and background)
& PNG (.ext)
\\
 {[}base{]}/subhalos/\{id\}/ \newline stellar\_mocks/sed.txt
& download stellar mock integrated 1D SED for subhalo \{id\}
& txt,json (.ext) \\ \hline
\multicolumn{3}{c}{\textbf{direct file downloads}}
\\ 
\multicolumn{3}{c}{define {[}base{]} = /api/{sim\_name}/files}
\\ 
\hline{[}base{]}/
& list of each 'files' type available for this simulation (excluding those attached to specific snapshots)
& json,api (?format=)
\\ 
 {[}base{]}/snapshot-\{num\}/
& list of all the actual file chunks to download snapshot \{num\}
& json,api (?format=)
\\ 
 {[}base{]}/snapshot-\{num\}.\{chunknum\}.hdf5
& download chunk \{chunknum\} of snapshot \{num\}
& HDF5 (.ext)
\\ 
 {[}base{]}/snapshot-\{num\}.\{chunknum\}.hdf5? \newline \{cutout\_query\}
& download only \{cutout\_query\} of chunk \{chunknum\} of snapshot \{num\}
& HDF5 (.ext)
\\ 
 {[}base{]}/groupcat-\{num\}/
& list of all the actual file chunks to download group catalog (fof/subfind) for snapshot \{num\}
& json,api (?format=)
\\ 
 {[}base{]}/groupcat-\{num\}.\{chunknum\}.hdf5
& download chunk \{chunknum\} of group catalog for snapshot \{num\}
& HDF5 (.ext)
\\
 {[}base{]}/lhalotree/
& list of all the actual file chunks to download {\sc LHaloTree} merger tree for this simulation
& json,api (?format=)
\\
 {[}base{]}/lhalotree.\{chunknum\}.hdf5
& download chunk \{chunknum\} of {\sc LHaloTree} merger tree for this simulation
& HDF5 (.ext)
\\
 {[}base{]}/sublink/
& list of all the actual file chunks to download {\sc SubLink} merger tree for this simulation
& json,api (?format=)
\\
 {[}base{]}/sublink.\{chunknum\}.hdf5
& download chunk \{chunknum\} of {\sc SubLink} merger tree for this simulation
& HDF5 (.ext) \\ \hline
    \end{tabular}
  \end{center}
\end{table*}

\end{document}